\documentclass[onecolumn,floatfix]{revtex4}
\usepackage{color}
\usepackage{epsfig}
\usepackage{graphicx}
\usepackage{amsmath}

\newcommand {\be}  {
\begin{equation}
}
\newcommand {\ee}  {
\end{equation}
}
\newcommand {\bea} {
\begin{eqnarray}
}
\newcommand {\eea} {
\end{eqnarray}
}

\newcommand {\m} {\mathcal}

\newcommand{\RN}[1]{%
  \textup{\uppercase\expandafter{\romannumeral#1}}%
}
\newcommand{\ignore}[1]{}

\newcommand{\boldnabla}{\text{\boldmath$\nabla$}}
\begin{document}

\title{Bi-phasic lithiation/delithiation dynamics in Li-ion batteries: application of the Smoothed Boundary Method in the phase field model}
\author{A. Yousfi$^{1,2}$, A. Demorti\`ere$^{1,2,3}$, G. Boussinot$^4$}
\affiliation{$^1$Laboratoire de Réactivité et de Chimie des solides (LRCS), Université de Picardie Jules Verne, CNRS UMR 7314, 33 rue Saint Leu, 80039 Amiens Cedex, France}
\affiliation{$^2$Réseau sur le stockage Electrochimique de l’Energie (RS2E), CNRS FR 3459, 33 rue Saint Leu, 80039 Amiens Cedex, France}
\affiliation{$^3$ALISTORE-ERI, FR CNRS, 80039, 3104, France}
\affiliation{$^4$Access e.V., Intzestr. 5, 52072 Aachen, Germany}

\begin{abstract}

An appropriate description of the lithiation/delithiation dynamics in bi-phasic primary cathode particles of Li-ion batteries requires an accurate treatment of the conditions holding at the interface between the particle and the surrounding liquid electrolyte. 
We propose a phase field model based on the Allen-Cahn approach within which the particle-electrolyte interface is smooth (Smoothed Boundary Method - SBM), in order to simulate arbitrarily shaped particles. 
Surface terms are added to the evolution equations, and SBM calculations are compared with benchmark simulations for which the boundary conditions are explicitly imposed at the borders of the calculation domain. 
We find that strengthening factors for the surface terms are needed in order to achieve the desired conditions for the phase and elastic fields, and to correctly reproduce the level of stress within the particle. 
This reveals crucial in the context of Li-ion batteries for an accurate prediction of the Li insertion/extraction rate and the Li diffusion behavior.
We perform also a simulation under potentio-static conditions with a full coupling of the different physical processes at play. It illustrates the applicability of our approach and demonstrates the capabilities of the SBM for a simulation of lithiation/delithiation dynamics with coupled electrochemistry and mechanics.

%Li-ion batteries are widely used as storage devices for electrical energy. Their lithiation/delithiation dynamics is associated with a phase transformation involving a Li-poor and a Li-rich phases, within which interstitial Li diffuses.  In so-called LFP batteries, the basic principle defines a physics in which the interstitial diffusion of Li within the host crystal is associated with a phase transformation between the Li-poor FP and Li-rich LFP phases, generating coherency stresses.     Here, we propose a phase-field model for the study of primary particles surrounded by a liquid electrolyte. We use the Allen-Cahn approach that allows for a consistent description of diffusive phase transformations. Moreover, we use the Smoothed Boundary Method to tackle the problem of reproducing desired conditions at the crystal/electrolyte interface. We show how their intricate nature renders the model design subtle.  

%
\end{abstract}

\maketitle

\section{Introduction}

Li-ion batteries are widely used as an efficient tool to store energy owing to their low cost, low environmental impact and long life-time. Their microstructure and heterogeneities are fundamental factors driving their electrochemical performances, mechanical properties, life-time and safety \cite{LBF2020}. Thus, studying the development of the microstructure is the crucial first step in order to find optimization routes leading to enhanced performances of the battery.

The lithiation/delithiation of a Li-ion batteries works under the principle of insertion/extraction of Li into/from a host crystal that constitutes the cathode particle. This alters its local chemical and electronic properties, as well as its lattice structure. For example, in LFP batteries, Li atoms are extracted from a FePO$_4$ host crystal during the charge, and are inserted back into it during the discharge. The extraction, associated with the loss of one electron, is driven by an applied electrical current or voltage. Then Li diffuses as Li$^+$ ions in the surrounding liquid electrolyte. Conversely, when Li inserts back, it gains an electron and then diffuses as an atom in the host crystal. This insertion generates the electrical current that allows for the usage of the electrical device. 

Typically, a Li-poor (FP) and a Li-rich (LFP) phases coexist thermodynamically at the operating  temperatures in such batteries \cite{biphasage1, biphasage2}. In the cathode particle, Li atoms diffuse as interstitials within the FP phase and Li-vacancies diffuse within the LFP phase. In addition, a phase transformation takes place when the interfaces between FP and LFP move. Across these interfaces, the Li occupancy varies abruptly, and their motion implies Li diffusion fluxes. The lithiation/delithiation dynamics thus corresponds to a phase transformation coupled to a diffusion process. This results in a complex pattern formation phenomenon implying various sources of Li fluxes and several length and time scales.

Phase field models have proven their capabilities in addressing such problems \cite{review_PF}. 
They describe continuous fields that obey the same evolution equations everywhere in the simulation domain, thereby avoiding an explicit tracking of the interfaces. 
They have been very widely applied, for example to dendritic \cite{karma_rappel, kai_dendrite} and eutectic \cite{folch, kai_eutectoid} growths, to stress-driven solid-solid transformations \cite{lq_chen, moi_nial}, or to two-fluids systems \cite{anderson}.
 
For Li-ion batteries, the most used approach until now relies on the Cahn-Hilliard approach. Originally, it describes the spinodal decomposition, for which small variations of the Li occupancy amplify due to a linear instability, and corresponds to a second-order phase transition. Within such models, a single field $C$ plays the role of a Li occupancy indicator, but also of an order parameter indicating whether the system is locally in the FP or LFP state \cite{gogi, coherent_nucleation}. The field $C$ is locally conserved and obeys a continuity equation involving the Li diffusion flux. 
Interesting results were obtained, for example the evidence of the two limiting surface reaction- and bulk diffusion-controlled regimes \cite{gogi}.

In classical out-of-equilibrium thermodynamics, the diffusion flux is assumed to result from the spatial variations of the Li chemical potential (it may also depend on variations of the chemical potential of the host crystal when cross diffusion occurs). The latter represents the local state of the system after coarse graining of the microscopic degrees of freedom and includes only bulk properties. 
However, within the Cahn-Hilliard approach, the Li chemical potential itself incorporates spatial variations of $C$. This is problematic if one requires a definition in line with classical out-of-equilibrium thermodynamics.
Moreover, following the phenomenology of first-order phase transitions, one may expect that more than one field is required to describe the system consistently. 
To support this statement, one should recall that two chemical potentials (one for the host crystal and one for the Li atoms) should be accounted for.
The Li flux across a moving FP-LFP interface is a quantity that is independent from the velocity of the interface (that represents the flux of host atoms across the interface), both of them responding to kinetic boundary conditions (BCs) involving the two chemical potentials \cite{caroli}. 

Within the so-called Allen-Cahn approach that we use here, this possibility is offered by the coupled evolution of a conserved field and a non-conserved field. While the conserved field describes the spatial evolution of the Li chemical potential (in an interstitial alloy the diffusion potential is the chemical potential of the interstitial atoms, while the diffusion potential is the difference between the chemical potentials of the solute and the solvent in a substitutional alloy), the non-conserved field describes the motion of the FP-LFP interface. Such a model was extensively used in metallurgy, for which close-to-equilibrium first-order phase transitions with complex growth phenomena commonly take place \cite{karma_rappel, folch}. Its classical formulation was extended to account for a kinetic cross-coupling between the two fields \cite{moi_cross_coupling}. 

\begin{figure}[t!]
  %\centering
  \includegraphics[keepaspectratio, width=0.85\textwidth]{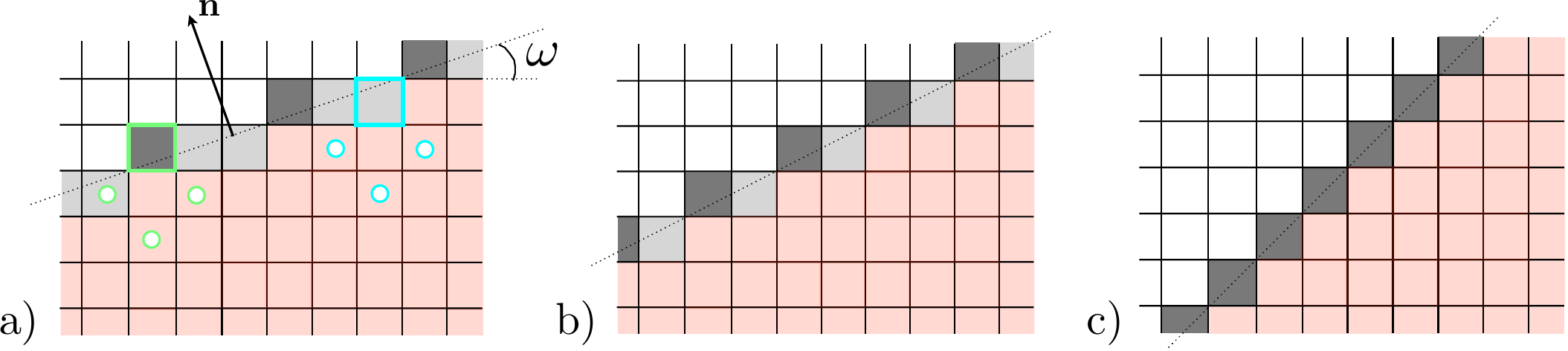}
  \caption{ Scheme showing, in (a), how boundary conditions (BCs) may be imposed at a sharp particle-electrolyte interface making an angle $\omega$ with the discretization grid. At the grid points in red, belonging to the physical domain (i.e. the particle), the field obeys its evolution equation. At the light and dark gray points, the field value is set so as to impose, in accordance with the orientation of the normal vector $\bf n$, targeted BCs such as von Neumann-type ones. This involves the field values designated by their respective white dots. While the light gray points (highlighted in blue) require only values from the physical domain, dark gray points (highlighted in green) require the value from a light gray one, suggesting to primarily set light gray points and secondarily set the dark gray ones. In (b), is illustrated the fact that when $\omega$ increases, the number of light gray points diminishes. In (c), $\omega=45^\circ$ and no light gray point exist, thereby hindering the possibility to properly set the BCs. }
  \label{scheme_bc}
\end{figure}

At the particle-electrolyte (P-E) interface, it is crucial for the phase field model to reproduce some physical laws, for example the electro-chemical reaction rate that describes the insertion/extraction of Li or the traction-free conditions for the elastic field that hold if the particle is surrounded by the electrolytic liquid. 
This provides BCs that have to be fulfilled by the theoretical formulation. 
Here, we aim at simulating the lithiation/delithiation dynamics for an arbitrary shape of the particle. 
We thus need a versatile procedure to impose BCs.
Commonly, when the evolution equation of a continuous field incorporates its spatial  derivatives, one defines a calculation domain, discretized with help of a grid, that contains additional grid points at which the value of the field is explicitly imposed so as to fulfill the targeted BC. 
The necessary number of additional points depends on the order of the partial derivative and on how its discretization is performed. 
When, as here, the physics is governed by bulk diffusion with second order derivatives, that concern only first neighbors when discretized using usual schemes, a single additional point is required at each position along the interface. 
Then, when a von Neumann-type BC is imposed (as it will be the case for the electro-chemical reaction for example), the value of the field at this additional point is fixed so that the scalar product of the gradient of the field and the vector normal to the interface  equals some targeted value. 
This is straightforward for an interface aligned with the axes of the discretized coordinate system and the same procedure is applied at each additional point. 
The situation is more complicated when it is not aligned. 
When the interface presents an angle $\omega$ with those axes, then two kinds of additional points exist, as seen in Fig. \ref{scheme_bc}a where they are designated by a light and dark gray color, while the grid points that belong to the physical domain, i.e. the particle, are in red. 
At the latter, the field obeys its evolution equation, and the von Neumann-type BC is achieved by imposing,  according to the orientation of the normal vector $\bf n$, the field value at the light and dark gray grid points (highlighted using the green and blue frames), using the field values at their respective white dots (we assume here a usual central finite difference scheme). 
We see that, while the light gray point necessitates only field values from within the particle, i.e. from red points, the dark gray point necessitates the field value from a light gray point.   
One way is thus to fix first the field value of the light gray points and then to fix the field value of the dark gray points.
However, we see from Fig. \ref{scheme_bc}b and \ref{scheme_bc}c that the number of light gray points diminishes when the angle $\omega$ increases, and that for $\omega = 45^\circ$, they are absent.
Within this frame with a sharp P-E interface, a proper imposition of BCs for a particle with arbitrary shape is thus hindered.  

Another theoretical tool to solve partial differential equations is the finite-element method \cite{shahed}. Within this frame, BCs are more easily imposed because they appear naturally in the so-called weak form of the evolution equations. However, the discretization of the space is not regular, which brings practical complications and inhibits the usage of simple parallelization schemes. 
Therefore, we use an alternative route.    

We  base our approach on the work in Ref. \cite{thornton}, in which the authors introduce the so-called Smoothed Boundary Method (SBM). An smoothly varying indicator field is used to distinguish the particle from the electrolyte, and appropriate surface terms are introduced in the evolution equations in order to achieve the targeted BCs at the P-E interface. The same concept was used in various manner, see \cite{levine, voigt}.

In this article, we want to demonstrate how the SBM applies to our problem of lithiation/delithiation dynamics. 
We  start by presenting the phase field model within the SBM formulation. 
We  then present its ability to achieve the different types of targeted BCs, specific to the lithiation/delithiation problem. 
For each of them, we  perform calculations in which we inhibit the physical processes that do not play a role in the establishment of the BC.
We  finally apply our model to a situation where all physical processes are at play. 
%with a full coupling of the three physical processes
%A one-dimensional diffusion problem under both monophasic and biphasic situations will be solved. Then, we will study the contact angle where the FP-LFP interface meets the P-E interface. We will also exhibit how SBM enables a mechanically traction-free P-E interface. {\color{blue}We will finally comment our results and evidence subtleties arising due to the intricate nature of the conditions that are imposed at the P-E interface.}           

\section{Phase Field model}\label{PF_model}

At coexistence of the FP and LFP phases at a given temperature (that we consider constant in time and homogeneous in space), the equilibrium Li concentration in the FP phase is $c^{FP}_{eq}$, and is $c^{LFP}_{eq}$ in the LFP phase ($1-c^{LFP}_{eq}$ is then the vacancy concentration). 
When the system is out of equilibrium, diffusion takes place and $c$ is the spatially varying local Li concentration. 
We then define a rescaled concentration field $C = (c - c^{FP}_{eq})/(c^{LFP}_{eq} - c^{FP}_{eq})$, whose value at coexistence is $C=1$ in LFP and $C=0$ in FP.

For our phase field model, we use the so-called Allen-Cahn approach. As mentioned in the introduction, it is especially suited for first order phase transformations. 
The biphasic evolution is described by a phase field $\phi$ whose spatial variations, localized on a length scale $W$, indicate the interface between the two phases FP and LFP. 
In the bulk of FP phase, we have $\phi=0$, and in the bulk of LFP phase we have $\phi=1$. 
In this work, we are interested by the coupled evolutions of $\phi$ and $C$ within the particle. This evolution is driven by chemical effects as well as by elastic effects arising due to a dependence on $C$ of the lattice parameters of the host crystal. 

One should note that the interface width $W$ is a purely numerical quantity. While it is in the order of the atomic distance in reality, phase field models are written in such a way that $W$ may be chosen order of magnitude larger. This is beneficial in terms of time consumption of the calculation because the interface needs to be resolved with only few grid points and one may thus simulate larger systems when increasing the interface width. However, the interface width should remain the smallest characteristic length scale of the fields' variations during the simulation. For example, if the interface is curved, its radius of curvature should remain much larger than $W$. This constraint comes from the design of the phase field model as a mathematical formulation for solving sharp-interface equations, where the interface is infinitely thin. Since $W$ is actually non-vanishing, spurious effects may arise and the so-called "thin-interface" analysis of the phase field model \cite{karma_rappel, moi_mbe} is then necessary to derive the interface BCs that the model effectively reproduces.

Within the particle, we use a free energy functional $F$:
\bea
F = \int_V f[\phi,C] dV
\eea
where the density $f$ is
\bea\label{free_energy_density}
f = H \left[\phi^2 (1-\phi)^2 + \frac{W^2}{2} |\boldnabla \phi|^2 \right] + \frac{X}{2} \left[ C - C_{eq}(\phi)\right]^2 + \frac{\lambda_{ijkl}}{2} \left[ \varepsilon_{ij} - \varepsilon_{ij}^0(C) \right] \left[ \varepsilon_{kl} - \varepsilon_{kl}^0(C) \right]. 
\eea
The term proportional to $H$ defines the equilibrium distribution of the phase field
\bea\label{eq_phi}
\phi_{eq}(z) = \frac{1}{2} \left[ 1 + \tanh\left( \frac{z}{\sqrt{2} W} \right)  \right] \label{eq_profile}
\eea
across an equilibrium and thus flat interface with a normal direction $z$, and with the LFP phase at $z \to \infty$ and the FP phase at $z \to -\infty$. 
The second term proportional to $X$ defines the free energy cost for a deviation from the $\phi$-dependent equilibrium concentration 
\bea
C_{eq}(\phi) = \phi^3(10-15 \phi+6\phi^2),
\eea
that equals $C_{eq}(\phi=1)=1$ and $C_{eq}(\phi=0)=0$ in accordance with the definition of the rescaled concentration $C$ given above. Here we use simple parabolas with the same curvature. One may instead use regular solution models \cite{gogi, nestler}. 
The derivative of $C_{eq}$ with respect to $\phi$, which we use later on, is $dC_{eq}/d\phi = 30 \phi^2 (1-\phi)^2$.
The third term in Eq. (\ref{free_energy_density}), for which an implicit summation over repeated indices is assumed, corresponds to the elastic energy due to the deformation 
\bea
\varepsilon_{ij} = \frac{1}{2} \left( \partial_i u_j + \partial_j u_i\right) 
\eea
of the crystalline lattice, with $u_i$ the displacement with respect to an arbitrary reference state. With respect to the same reference state, the crystalline lattice possesses an equilibrium deformation depending on $C$ via a linear Vegard law: 
\bea \label{eps0}
\varepsilon^0_{ij}(C) = \varepsilon_{ij}^{FP} + \left[ \varepsilon_{ij}^{LFP} - \varepsilon_{ij}^{FP}\right] C .
\eea  
This provides $\varepsilon^0_{ij}(C=0) = \varepsilon_{ij}^{FP}$ and $\varepsilon^0_{ij}(C=1) = \varepsilon_{ij}^{LFP}$ as stress-free deformation in equilibrium FP and LFP phases respectively. 
Due to the Li intercalation in the host crystal, which stretches the crystal unit cell, we have $\varepsilon_{ij}^{LFP}>\varepsilon_{ij}^{FP}$. 
The elastic constants $\lambda_{ijkl}$ are assumed to be phase and concentration independent, i.e. independent of $\phi$ and $C$, for simplicity.  
\\

In practice, prescribing the ratio between the FP-LFP interface width $W$ and the associated capillary length fixes the energy scale $H$ ($HW$ is the FP-LFP interface energy, to which is proportional the capillary length). Then one fixes the ratio of $H$ and $\lambda_{ijkl}$ by comparing to $W$ the length scale provided by the ratio of the FP-LFP interface energy (that can probably be very anisotropic with low values such as 10$^{-2}$J/m$^2$ \cite{thornton2}) and the elastic constants (in the order of 100GPa). 
And without loss of generality, we may consider  from now on a dimensionless free energy density and set $X=1$. 
\\

\begin{figure}[t!]
  %\centering
  \includegraphics[keepaspectratio, width=0.85\textwidth]{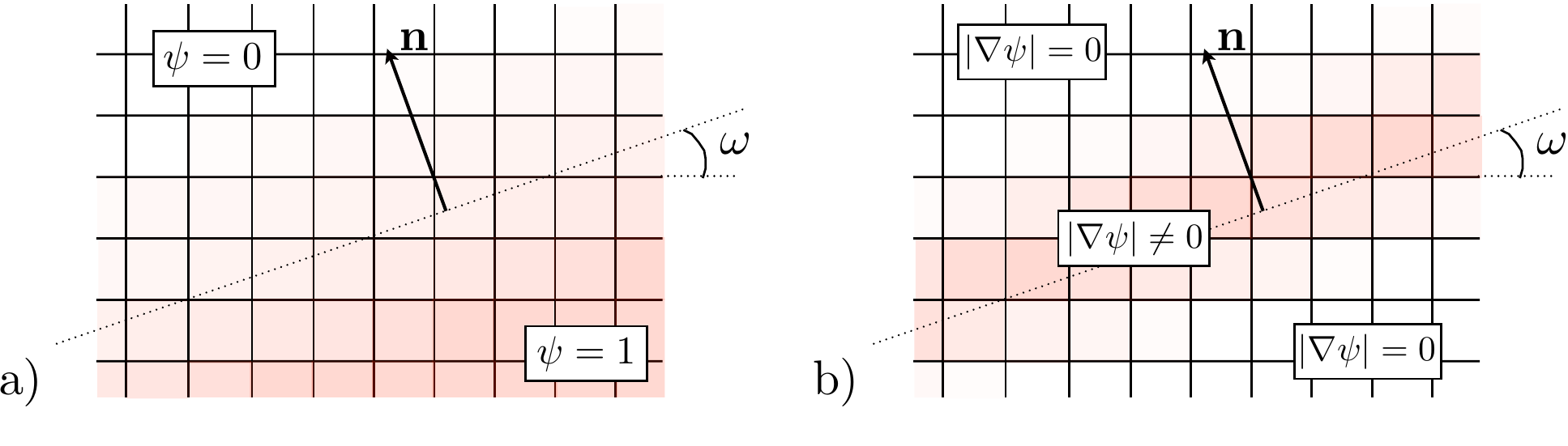}
  \caption{ Scheme illustrating the SBM definition of the smoothly varying indicator field $\psi$. In (a), the color indicates the magnitude of $\psi$, and in (b) it indicates the magnitude of its gradient $|\boldnabla \psi |$. The normal vector, given by ${\bf n} = \boldnabla \psi / |\boldnabla \psi|$, is used to define the surface terms characterizing  the SBM [see Eqs. (\ref{dot_phi_surface}), (\ref{dot_mu_surface}) and (\ref{force})]. }
  \label{scheme_sbm}
\end{figure}

Moreover, we aim at developing a model that allows us to reproduce appropriate conditions at the boundary between the particle and the electrolyte. For this, we use the so-called Smoothed Boundary Method (SBM) \cite{thornton}. In a similar way as the phase field $\phi$, a domain field $\psi$ distinguishes between the particle P ($\psi=1$) and the electrolyte E ($\psi=0$) as illustrated in Fig. \ref{scheme_sbm}a. 

During the simulation of the lithiation/delithiation dynamics, this field remains constant and determines the shape of the cathode particle, i.e. the shape of the P-E interface. 
It obeys the equilibrium equation (\ref{eq_phi}), and its derivative  
\bea \label{psi_prime}
\psi'(z) = \frac{1}{2\sqrt{2} W} \left[ 1 - \tanh^2\left( \frac{z}{\sqrt{2} W} \right)  \right] 
\eea
verifies 
\bea\label{identity}
\psi'^2(z) = \frac{2}{W^2} \psi^2(1-\psi)^2 ,
\eea
 and is non-vanishing only in a region of width of order $W$, i.e. only at the P-E interface, as illustrated in Fig. \ref{scheme_sbm}b. 
This property is used in the SBM approach to localize at the P-E interface appropriate contributions to the conserved and the non-conserved equations. 
\\

We work with the variable
\bea \label{def_mu}
\mu =C - C_{eq}(\phi).
\eea
Using Eqs. (\ref{free_energy_density}) and (\ref{eps0}), we then have
\bea \label{def_df_dc}
\frac{\delta F}{\delta C} = \mu - \sigma_{ij} \varepsilon^*_{ij}.
\eea
where
\bea \label{stress}
\sigma_{ij} &=&  \frac{\delta F}{\delta \varepsilon_{ij}} = \lambda_{ijkl} \left[ \varepsilon_{kl} - \varepsilon_{kl}^0(C) \right] , \\
\varepsilon^*_{ij} &=& \varepsilon_{ij}^{LFP} - \varepsilon_{ij}^{FP} >0.
\eea
We now present the evolution equations, which are classical equations supplemented by SBM surface terms. 
\\

\subsection{Phase field equation}  \label{pf_eq}
The evolution of the phase field $\phi$ describes the motion of the FP-LFP interface. When the latter intersects the P-E interface, a balance of surface tensions imposes a contact angle that we denote $\theta_0$. This angle is imposed in the phase field model with the help of the SBM and the evolution equation for $\phi$ then reads:
\bea \label{dot_phi}
 \dot \phi =\frac{1}{\tau} \left( \Phi_{\text{bulk}} + \Phi_{\text{surface}} \right)
\eea
where
\bea
\Phi_{\text{bulk}} &=& - \frac{\delta F}{\delta \phi} = 30 \phi^2 (1-\phi)^2 \mu + H \left[ -2\phi(1-3\phi+2\phi^2) + W^2 \boldnabla^2 \phi  \right], \label{dot_phi_bulk}  \\
\Phi_{\text{surface}} &=& \beta_\phi H W \sqrt{2} (1-\psi) \left( \boldnabla \phi \cdot {\bf n}   + |\boldnabla \phi| \cos \theta_0 \right), \label{dot_phi_surface} \\
{\bf n} &=& \frac{\boldnabla \psi}{| \boldnabla \psi |}.
\eea
The unit vector $\bf n$ defines the normal direction to the P-E interface (see Fig. \ref{scheme_sbm}).
We should note that, using Eq. (\ref{identity}), we can rewrite the surface term as $\Phi_{\text{surface}} = \beta_\phi W^2 (\boldnabla \phi \cdot \boldnabla \psi + |\boldnabla \phi | |\boldnabla \psi| \cos \theta_0)/\psi$. The division by $\psi$ comes from the strategy that is used to derive the SBM equations (see Ref. \cite{thornton}), which explicitly incorporate the targeted BCs at the P-E interface. 
The vanishing of the term in parentheses corresponds to a contact angle $\theta_0$ between the FP-LFP interface (where $|\boldnabla \phi | \neq 0$)  and the P-E interface (where $|\boldnabla \psi | \neq 0$), which is the condition we want to fulfill.

Here, we have introduced the coefficient $\beta_\phi$, which is not present when the SBM equations are derived as in Ref. \cite{thornton} (then $\beta_\phi=1$). The reason for this introduction is that, as we will see later, it allows for a better accuracy of the SBM when compared to simulations with a direct imposition of BCs, that we will use as benchmarks. The goal being to achieve a vanishing of $\Phi_{\text{surface}}$, the coefficient $\beta_\phi$ prescribes the strength with which we want to impose our BCs. 
As can already be inferred from Eqs. (\ref{dot_phi_bulk}) and (\ref{dot_phi_surface}), the ability of our SBM to achieve a vanishing of the surface term will depend on the value of $\mu$. This question will be addressed in Section \ref{results_angle}.  
%We will see that increasing $\beta_\phi$, apart from possibly requiring a decrease in time discretization step and thus slowing down our calculation, may lead to larger errors, which imposes us to find an optimum.    
\\

\subsection{Diffusion equation} 
In the bulk of either phase, FP or LFP, the Li concentration $C$ in the crystalline lattice obeys the continuity equation
\bea
\dot C = - \boldnabla \cdot {\bf J} = -\partial_i J_i
\eea
where the $i$-th component  of the diffusion flux $\bf J$ is 
\bea
J_i = -D_{ij} \partial_j \frac{\delta F}{\delta C} 
\eea
with $D_{ij}$ the diffusion tensor. When considering it diagonal, i.e. $D_{ij} = \delta_{ik} \delta_{kj} D^0_k$ with $D_k^0$ the diffusion coefficient in the $k$-th direction ($\delta_{ik}$ is the Kronecker tensor), we have $J_x = D^0_x \partial_x (\delta F/\delta C)$ and the same for the other directions.

Noticing that, according to Eq. (\ref{def_mu}), $\dot C = \dot \mu + \dot C_{eq}(\phi)$, we arrive at the conservation equation for the Li atoms in terms of $\mu$ in the bulk of the particle:
\bea
\dot \mu|_{\text{bulk}} = -30 \phi^2(1-\phi)^2 \dot \phi + \partial_i \left(  D_{ij}  \partial_j \frac{\delta F}{\delta C} \right).
\eea
Within the SBM, the time derivative of $\mu$ is supplemented by a contribution from the surface of the particle, i.e. 
\bea
\dot \mu = \dot \mu|_{\text{bulk}} + \dot \mu|_{\text{surface}} \label{dot_mu}
\eea
with
\bea \label{dot_mu_surface}
\dot \mu|_{\text{surface}} = \frac{\sqrt{2} (1-\psi)}{W} \left(  - {\bf J} \cdot {\bf n} + \text{J}_{\text{ins}} \right).
\eea
Here also, $\bf n = \boldnabla \psi/|\boldnabla \psi|$.
 And again, a vanishing surface term corresponds to the desired BC. It ensures that the diffusion flux in the direction normal to the P-E interface ${\bf J} \cdot {\bf n} $ equals the electrochemical reaction rate $\text{J}_{\text{ins}}$. The latter may be obtained from text books on electrochemistry 
 %(see for example {\color{red}CITE}) 
 and is usually assumed to follow the Butler-Volmer form  
\bea \label{reaction_rate}
\text{J}_{\text{ins}} =  \text{J}_{\text{ins}}^0 \left[ \exp \left( \alpha \Delta \right) - \exp \left( -(1-\alpha) \Delta \right) \right] ,
 \eea
 where $\alpha$ is some symmetry factor of order unity, and where the velocity scale $\text{J}_{\text{ins}}^0$ is positive.
 A positive (negative) $\text{J}_{\text{ins}}$ corresponds to insertion (extraction).
 The reaction rate becomes linear, i.e. $\text{J}_{\text{ins}} =  \text{J}_{\text{ins}}^0 \Delta$, when $|\Delta| \ll 1$. Here, $\Delta$ is the so-called overpotential, to which is subjected the surface of the particle. 
 It describes, for $\Delta>0$, the driving force for a Li$^+$ ion in the electrolyte to be inserted, i.e. to cross the P-E interface and become an atom having gained an electron and diffusing in the crystal. For $\Delta<0$, it describes the driving force for a Li atom diffusing in the crystal to be extracted, i.e. to cross the P-E interface and become a Li$^+$ ion in the electrolyte, thereby leaving a mobile electron in the crystal. 

 The overpotential $\Delta$  incorporates the difference of electrostatic potential between the electrolyte and the cathode particle.
 %, that we assume independent of the distribution of Li within the particle and of Li in the electrolyte. 
 The overpotential also incorporates the so-called open-circuit potential,
 %, and when the latter exactly equals the electrostatic potential, the overpotential vanishes and neither insertion nor extraction of Li takes place. 
 %The open-circuit potential
 that depends on the state of the surface, i.e. its phase $\phi$ and its chemical potential $\delta F/\delta C$. 
 We  assume that $\Delta$ is defined as 
 \bea \label{Delta}
 \Delta = \mu_0 (\phi) - \frac{\delta F}{\delta C} .
 \eea
 According to Eq. (\ref{def_df_dc}), the reaction rate $\text{J}_{\text{ins}}$ decreases when $\mu$ increases, i.e. when the Li concentration increases. 
 Thus, insertion ($\text{J}_{\text{ins}}>0$) will be slowed down and extraction ($\text{J}_{\text{ins}}<0$) will be accelerated  by an increase in Li concentration.
 Moreover, $\text{J}_{\text{ins}}$ is also lowered (slowed down insertion and accelerated extraction) by a negative stress $\sigma_{ij}$, i.e. by a compression of the crystalline lattice, as expected since the presence of Li in the host crystal tends to stretch the lattice. 
 
 We may in general consider a $\phi$-dependent open-circuit potential if, for a given $\delta F/\delta C$, it differs between FP and LFP. 
 One may for example assume that 
 \bea
 \mu_0(\phi) = \mu_0^0 - \mu_0^{FP} - (\mu_0^{LFP} - \mu_0^{FP}) \phi.
 \eea
The electrochemical potential $\mu_0^0$ is the contribution from the electrostatic potential difference between the electrolyte and the particle, linked to it via the Faraday constant \cite{review_PF}. Then $\mu_0^{FP}$ $(\mu_0^{LFP})$ corresponds to the open-circuit potential of a purely FP (LFP) particle, i.e. $\phi=0$ $ (\phi=1)$, with a chemical potential $\delta F/\delta C=0$. 
  
In this work, we do not solve the Poisson equation for the electrostatic potential \cite{warren} and we assume that its jump across the P-E interface is independent of the Li distribution within the particle and the electrolyte. 
This allows us to consider that $\mu^0_0$ is constant in time when the experimental set up imposes a voltage difference between the two extremities of the apparatus (potentio-static process). 
Then the time-dependent electrical current $\text{I}(t)$ is given by
 \bea\label{galvanostatic}
\text{I}  =  \int_V   ({\bf J} \cdot \boldnabla \psi ) dV  = \int_V  \text{J}_{\text{ins}} |\boldnabla \psi| \; dV ,
 \eea
and the state of charge 
 \bea \label{state_of_charge}
 \langle C \rangle = \frac{\int_V \psi C \;dV }{\int_V \psi \; dV }
 \eea
 evolves according to 
 \bea \label{relation_i}
 \dot {\langle C \rangle} = \frac{\text{I}}{\int_V \psi \; dV } \;.
 \eea
 Conversely, the experimental set up may impose a constant electrical current (galvanostatic process) $\text{I}(t) = \text{I}_{\text{app}}$. Then ${\mu^0_0}$ becomes time-dependent, and one should update ${\mu^0_0}(t)$ accordingly at each time step. 
 In the general case, one should find the solution of a transcendental equation, i.e. the value of ${\mu^0_0}$ that satisfies Eq. (\ref{galvanostatic}) using a $\text{J}_{\text{ins}}$ given by Eq. (\ref{reaction_rate}).
However, within the linear approximation for $\text{J}_{\text{ins}}$, this value is explicitly given by
 \bea \label{mu00}
 \mu^0_0 = \mu_0^{FP}+\frac{1}{\int_V dV |\boldnabla \psi|} \left\{ \frac{\text{I}_{\text{app}}}{\text{J}^0_{\text{ins}}} + \int_V dV   \left[  \frac{\delta F}{\delta C} + \left(\mu_0^{LFP} - \mu_0^{FP} \right)  \phi \right] |\boldnabla \psi|\right\} .
 \eea 
 
 Note here that, in opposition to Section \ref{pf_eq}, no strengthening factor has been introduced in Eq. (\ref{dot_mu_surface}). We will come back to this point in Section \ref{results_angle}.
 
 \subsection{Elastic deformation} \label{model_elast} The SBM that is used for solving the elastic field within the particle relies on setting $\lambda_{ijkl} = \lambda_{ijkl}(\psi) =  \lambda_{ijkl}^0 \psi$ in Eqs. (\ref{free_energy_density}) and (\ref{stress}), where $\lambda_{ijkl}^0$ are the elastic constants of the crystal (we recall that we assume them homogeneous within the particle, i.e. independent of $\phi$ and $C$). 
 This definition enables to achieve a traction-free surface of the particle, as expected when the latter is isolated within the electrolytic liquid. 
 The elastic equilibrium, that we assume to establish on a short time scale compared to diffusion and phase transformation, establishes when the force $\mathcal F_i$, sum of a bulk and a surface term, vanishes, i.e. 
\bea \label{elast_eq_SBM}
\mathcal F_i = \mathcal F_i|_{\text{bulk}} + \mathcal F_i|_{\text{surface}} = 0.
\eea
The bulk term is the classical 
 \bea 
\mathcal F_i|_{\text{bulk}} =  \psi \lambda_{ijkl}^0 \partial_j [\varepsilon_{kl} - \varepsilon^0_{kl}(C)]
 \eea
 and the surface term reads
 \bea
\mathcal F_i|_{\text{surface}}=  \beta_{\text{el}}  \lambda_{ijkl}^0 (\partial_j \psi) [\varepsilon_{kl} - \varepsilon^0_{kl}(C)]. \label{force}
 \eea
 
The vanishing of the surface term ensures a traction-free surface of the particle, where  
\bea 
 \sigma_{ij}  (\partial_j \psi) = 0,
 \eea
 with $\sigma_{ij}$ given by Eq. (\ref{stress}).
 Here we see that we have also introduced a strengthening factor $\beta_{\text{el}}$. The latter is not present, i.e. $\beta_{\text{el}}=1$, if one writes the elastic equilibrium as:
\bea\label{elast_eq_classic}
\partial_j \sigma_{ij} = 0. 
\eea
As we will show in Section \ref{results_elast}, the vanishing of the surface term is rather poor when $\beta_{\text{el}}=1$ and much better when $\beta_{\text{el}}$ is larger.

\section{Tests simulations}

Let us now present the performances of the SBM when applied to our problem of a bi-phasic lithiation/delithiation dynamics influenced by elastic deformations. We have made three types of tests, each of which corresponds to the imposition of a certain type of BC at the P-E interface. First, we aim at reproducing an electrochemical reaction rate in accordance with the overpotential at the surface of the particle. For this we perform one-dimensional simulations of a strictly diffusive problem (mono-phasic) and of a problem where diffusion is coupled to the phase transformation (bi-phasic). Second, we perform two-dimensional simulations in which we aim at imposing a contact angle $\theta_0$ at the point where the FP-LFP and the P-E interface meet. Third, still in two dimensions, we aim at achieving a traction-free surface of the particle. Those tests consist in performing SBM calculations in an ideal geometry that allows to perform corresponding  calculations in which we impose explicitly the desired BCs. We are then able to benchmark our SBM formulation. In the following, we measure all lengths in units of $W$, i.e. we set $W=1$.

\subsection{Electrochemical reaction rate}

\subsubsection{Mono-phasic one-dimensional problem}

First, we consider a purely diffusive one-dimensional mono-phasic system with $\phi(x)=0$ (or equivalently $\phi(x)=1$) and thus $\dot \phi=0$. Elastic effects are here neglected so that $\delta F/\delta C = \mu$.
As a reaction rate, we want to impose a Li flux entering the particle that linearly depends on the chemical potential $\mu$ at the P-E interface, i.e. $\text{J}_{\text{ins}} = \text{J}_{\text{ins}}^0(\mu_0 - \mu)$. 

For SBM, we use a domain of length $L=60$, and we set $\psi(x)= (1/2) \left[1 - \tanh \left(x/\sqrt{2} \right) \right]$. 
The P-E interface where $\psi$ varies is thus located around $x=0$ $\left(-1 \lesssim x \lesssim 1 \right)$, the particle is located at $-30 \leq x \leq 0$ ($\psi \approx 1$ for $-30 \leq x \lesssim -1$) and the electrolyte at $0 \leq x \leq 30$ ($\psi \approx 0$ for $1 \lesssim x \leq 30$).
The condition that we aim at imposing at the P-E interface is the vanishing of $\dot \mu|_{\text{surface}}$ in  Eq. (\ref{dot_mu_surface}), i.e. $D\partial_x \mu = \text{J}_{\text{ins}}$ here in absence of elastic effects.
 We thus solve Eq. (\ref{dot_mu}) that reduces to :
\bea
\dot \mu = D \partial_{xx} \mu + \frac{\sqrt{2} (1-\psi) D}{W} \left(-\partial_x \mu + \frac{\mu_0 - \mu}{l_{\text{ins}}} \right) \;\;\; \text{for SBM.}
\eea 
Here, we keep $W$ in the equation so that the latter remains dimensionally consistent.
 The diffusion coefficient $D$ is assumed constant (same diffusivity for atoms and vacancies)  and the reaction rate is defined through the length scale $l_{\text{ins}}$, i.e. $\text{J}_{\text{ins}}^0 = D/l_{\text{ins}}$. Moreover, we impose $\partial_x \mu = 0$ at $x=-30$ thereby imposing a mirror symmetry to the particle. We impose also $\partial_x \mu = 0$ at $x=30$. However, this condition is irrelevant and another one could have been chosen since the dynamics in the region where $\psi=0$ is anyway irrelevant.

For the benchmark simulation, we use a domain length $L=30$ and the particle is located at $-30 \leq x \leq 0$. In this case, $\psi$ does not exist (or may be set to $\psi=1$). We thus solve 
\bea
\dot \mu = D \partial_{xx} \mu \;\;\; \text{for the benchmark,}
\eea 
and we impose $\partial_x \mu = 0$ at $x=-30$ (mirror symmetry) and the Li flux $\partial_x \mu = (\mu_0 - \mu)/l_{\text{ins}}$ at $x=0$.

\begin{figure}[t!]
  %\centering
  \includegraphics[keepaspectratio, width=0.5\textwidth]{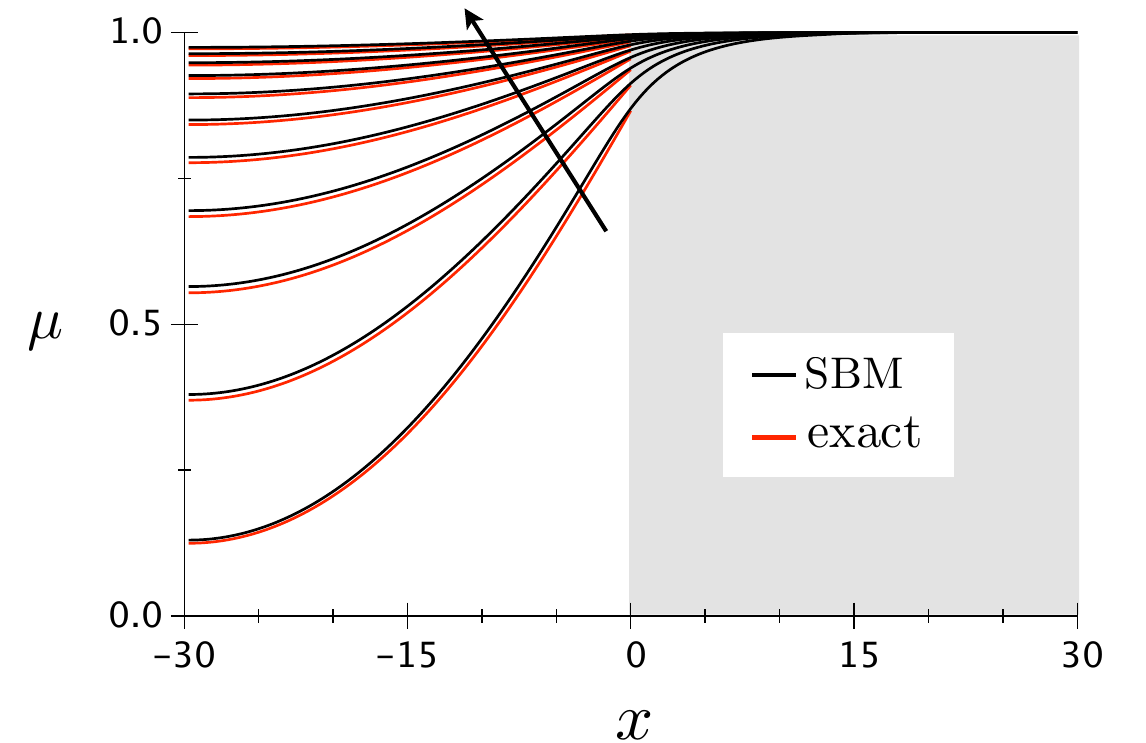}
  \caption{ Electrochemical reaction rate: mono-phasic 1D diffusion problem. Comparison, between SBM and the benchmark "exact" solution, of the profile of chemical potential $\mu$ in the physical domain $x<0$ where the indicator field $\psi=1$. Here $l_{\text{ins}}=3$. The gray region corresponds to the extension of the calculation domain where the indicator field vanishes, i.e. $\psi=0$. The arrow indicates the progression of time.}
  \label{chempot_monophase}
\end{figure}

\begin{figure}[t!]
  %\centering
  \includegraphics[keepaspectratio, width=0.7\textwidth]{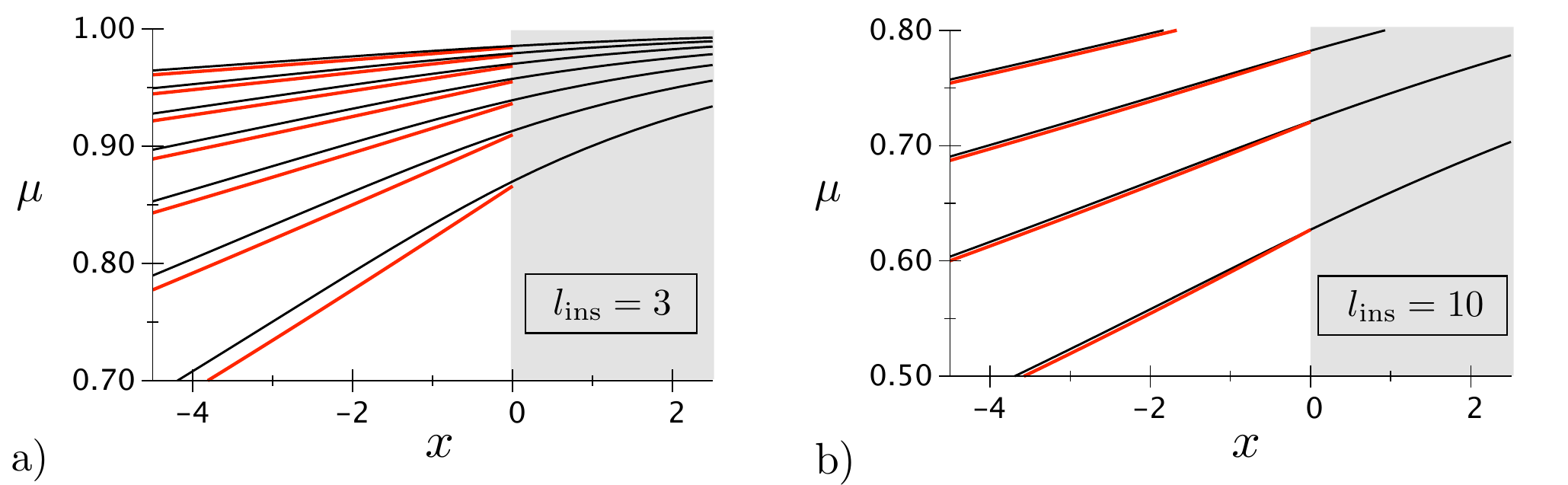}
  \caption{ (a) Close-up from Fig. \ref{chempot_monophase} where $l_{\text{ins}}=3$;  (b) Same for $l_{\text{ins}}=10$. We observe that the agreement between SBM and benchmark increases with $l_{\text{ins}}$.}
  \label{chempot_monophase_zoom}
\end{figure}

Initially, $\mu(x)=0$ for both SBM and benchmark simulations, and we set $\mu_0=1$ for convenience and without loss of generality (in this one-dimensional problem, $\mu_0$ is an arbitrary scale). 
In Fig. \ref{chempot_monophase}, we present for the same times the chemical potential $\mu$ for the SBM (black) and the benchmark (red) that is called "exact". Here $l_{\text{ins}}=3$ and the arrow indicates the course of time. The gray region corresponds to $\psi=0$. 
In the other region, i.e. the physical region where $\psi=1$, we observe a quasi-superposition of the profiles, alike  in Refs. \cite{levine, thornton}. 
This demonstrates that the SBM is able to precisely reproduce the targeted BC.
The latter neither corresponds to a purely von Neumann nor to a Dirichlet BC. While  $\partial_x \mu$ is prescribed for the former, the value of $\mu$ itself is prescribed for the latter.     
Here, the value of $\partial_x \mu$ is prescribed by $\mu$, which itself evolves according to an evolution equation that explicitly incorporates $\partial_x \mu$. 
The successful comparison that is presented in Fig. \ref{chempot_monophase} thus represents a more stringent test of the SBM than in \cite{levine, thornton}. 
Moreover, one should note that the agreement increases with $l_{\text{ins}}$. 
This is illustrated in Fig. \ref{chempot_monophase_zoom} where we zoom on the P-E interface for $l_{\text{ins}} = 3$ (a) and $l_{\text{ins}} = 10$ (b). 
% insertion length is $l_{\text{ins}} = 3$ in Fig. \ref{chempot_monophase}. In Fig. \ref{chempot_monophase_zoom}, we show using a zoom on the P-E interface that the agreement increases with $l_{\text{ins}}$ This is a rather small value being larger but not much larger than $W$, and one should note that when increasing to $l_{\text{ins}}/W=10$ and then $l_{\text{ins}}/W=100$, the agreement becomes better and better. 
\\

\subsubsection {Bi-phasic problem} \label{biphasic}

We now compare the SBM and the exact solution in the case where diffusion is coupled to the FP-LFP phase transformation, i.e. in the bi-phasic case for which $\dot \phi \neq 0$. 
We want to impose the same reaction rate as in the mono-phasic case.

For the SBM, the domain length is $L=180$ and we set $\psi(x)=(1/2)\left[ 1 - \tanh \left( x/\sqrt{2} \right) \right]$. Thus the particle is located at $-120 \leq x \leq 0$, the electrolyte at $0 \leq x \leq 60$, and the P-E interface located around $x=0$. 
For the benchmark, the simulation domain is  $-120 \leq x \leq 0$.
Moreover, the phase field is initialized, for SBM and the benchmark, as $\phi(x) = (1/2) \left[ 1+\tanh \left( (x+18)/\sqrt{2} \right) \right]$, i.e. with the FP phase ($\phi=0$) at $-120 \leq x \leq -18$ and the LFP phase ($\phi=1$) at $-18 \leq x \leq 0$. 

Here $\dot \phi \neq 0$ and  we thus solve for the chemical potential 
\bea
\dot \mu = -30 \phi^2(1-\phi)^2 \dot \phi + D \partial_{xx} \mu + \frac{\sqrt{2} (1-\psi) D}{W} \left(-\partial_x \mu + \frac{\mu_0 - \mu}{l_{\text{ins}}} \right) \;\;\; \text{for SBM,}
\eea
with $\partial_x \mu=0$ at $x=-120$ (mirror symmetry) and $x=60$ (irrelevant BC), and
\bea
\dot \mu = -30 \phi^2(1-\phi)^2 \dot \phi + D \partial_{xx} \mu  \;\;\; \text{for the benchmark,}
\eea
with $\partial_x \mu = 0$ at $x=-120$ and the Li flux $\partial_x \mu = (\mu_0 - \mu)/l_{\text{ins}}$ at $x=0$. 
The diffusion coefficient $D$ is constant as in the mono-phasic case.

In one-dimension, the surface term $\Phi_{\text{surface}}$ introduced in Eq. (\ref{dot_phi_surface}) to prescribe the angle made by the FP-LFP and P-E interfaces does not have any meaning. Thus we solve 
\bea
\dot \phi = \frac{1}{\tau} \left\{ 30 \phi^2(1-\phi)^2 \mu + H \left[ -2 \phi(1 - 3 \phi + 2 \phi^2) + W^2 \partial_{xx} \phi \right] \right\} \;\;\; \text{for both SBM and  benchmark,}
\eea
and we impose zero-slope von Neumann BCs, i.e. $\partial_x \phi (x=-120) = \partial_x \phi (x=60) = 0$ for SBM and $\partial_x \phi (x=-120) = \partial_x \phi (x=0) = 0$ for the benchmark.

We take $H=1$ and we choose the mobility time scale $\tau$ in accordance with the "thin-interface" limit mentioned in the Introduction \cite{karma_rappel, moi_mbe}. Then $\tau = (\chi/ \xi) W^2/(4 D)$ where $\chi = 4 \int_{-\infty}^\infty  q[\phi_{eq}(z)](1-q[\phi_{eq}(z)]) (dz/W) \approx 1.40748$ with $q(\phi)=\phi^3(10 - 15 \phi + 6\phi^2)$ and $\phi_{eq}(z)$ given by Eq. (\ref{eq_phi}), and $\xi = W \int_{-\infty}^\infty [\phi'_{eq}(z)]^2 dz = \sqrt{2}/6$ with $\phi_{eq}'(z)$ given by Eq. (\ref{psi_prime}). 

 Here also the chemical potential is $\mu(x)=0$ initially, and $\mu_0=1$. We choose $l_{\text{ins}}=10$.
 
\begin{figure}[t!]
  %\centering
  \includegraphics[keepaspectratio, width=0.9\textwidth]{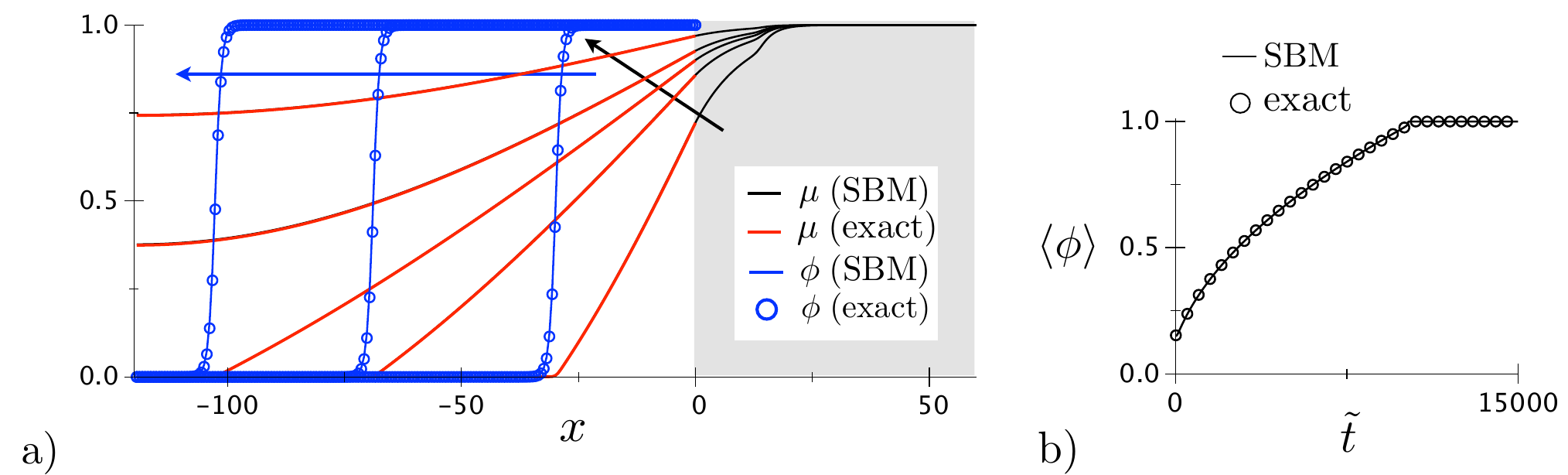}
  \caption{ Electrochemical reaction rate: bi-phasic 1D diffusion problem. Here $l_{\text{ins}}=10$. Comparison, between SBM and the benchmark "exact" solution, for: (a) the profile of chemical potential $\mu$ and of phase field $\phi$; (b) the evolution as a function of $\tilde t=Dt/W^2$ of $\langle \phi \rangle$ (see text for definition). 
 In (a), the gray region corresponds to $\psi=0$, and the arrows indicate the progression of time.}
  \label{chempot_biphase}
\end{figure}
 In Fig. \ref{chempot_biphase}a, we present for 5 equivalent times the chemical potential $\mu$ from the SBM in black color and for the "exact" solution (benchmark) in red color. The black arrow indicates the course of time. We observe a quasi-perfect superposition of the profiles. 
 In addition we present in blue color the phase field $\phi$ for the three first times for which $\mu$ is shown ($\phi(x)=1$ for further times). The SBM solution is in solid line, and the "exact" one is with circle symbols. We also observe a perfect coincidence of the profiles. 
 More globally, we see in Fig. \ref{chempot_biphase}b that the phase fraction $\langle \phi \rangle = (\int_x  \phi \psi \;dx)/ (\int_x \psi \;dx)$  for SBM and $\langle \phi \rangle = (1/L) \int_x  \phi \;dx$ for the benchmark are undistinguishable over the whole simulation time $\tilde t = Dt/W^2$.

To conclude on the 1D diffusive simulations, the SBM is very efficient to correctly reproduce, even when coupled to the phase field $\phi$, the spatio-temporal evolution of the chemical potential $\mu$ subjected to the prescribed electrochemical insertion/extraction rate at the P-E interface. 
\\

\subsection {Contact angle at the triple junction} \label{results_angle}

\begin{figure}[t!]
  %\centering
  \includegraphics[keepaspectratio, width=0.7\textwidth]{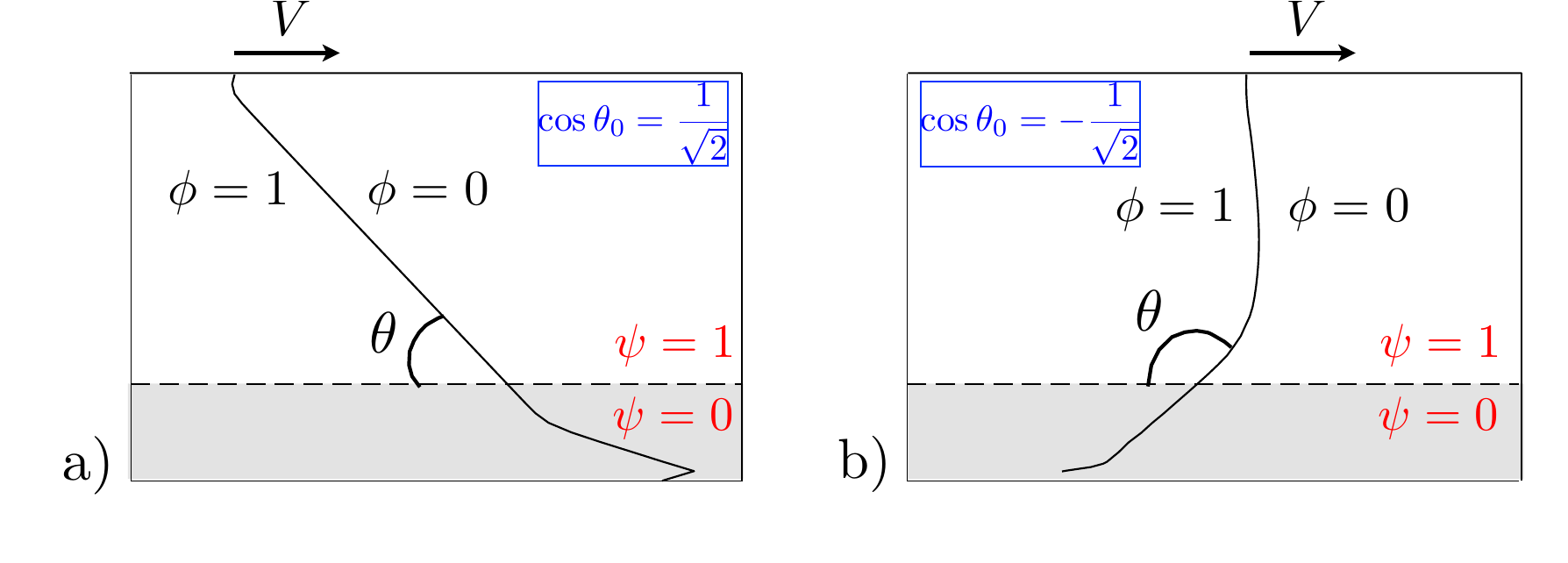}
  \caption{Contact angle at the triple junction. Location of the FP-LFP interface, i.e. locus of $\phi = 0.5$, moving with velocity $V$,  in the case where: (a) $\theta_0 = \pi/4$ and $\cos \theta_0 = 1/\sqrt{2}$; (b) $\theta_0 = 3\pi/4$ and $\cos \theta_0 = -1/\sqrt{2}$. The FP-LFP interface intersects the P-E interface with an angle $\theta$ that approaches $\theta_0$ owing to the surface term in Eq. (\ref{dot_phi}).}
  \label{contact_angle_schema}
\end{figure}

Let us now investigate the ability of the SBM to correctly prescribe the contact angle $\theta_0$ at the position where the FP-LFP and the P-E interfaces meet.
 Here, we do not need benchmark calculations and we simply measure the angle $\theta$ that results from the SBM calculations and compare it with $\theta_0$. 
We perform two-dimensional simulations without elastic effects. For simplicity, we set a constant in time and homogeneous in space chemical potential $\mu=\mu_0$.  
Thus we do not solve Eq. (\ref{dot_mu}), and we solve Eq. (\ref{dot_phi}) with $\mu=\mu_0$ in Eq. (\ref{dot_phi_bulk}). 
Moreover, the phase field time scale $\tau$ becomes arbitrary and simply defines the mobility of the interface (in comparison, when the evolution of $\phi$ is coupled to the diffusive process with a spatially varying $\mu$, the local equilibrium condition at the FP-LFP interface imposes the choice that was made in Section \ref{biphasic} for $\tau$ according to the thin-interface limit).
Owing to the absence of diffusion, a quasi steady-state is almost immediately found by the system and we present in Fig. \ref{contact_angle_schema} the corresponding pattern (the black solid line denotes the locus $\phi=0.5$) with the measured angle $\theta$ for the two types of calculation that we perform, with $\theta_0 = \pi/4$, i.e. $\cos \theta_0 = 1/\sqrt{2}$, and with $\theta_0 = 3\pi/4$, i.e. $\cos \theta_0 = -1/\sqrt{2}$. 
Note that at the upper boundary of the simulation domain, a no-slope condition is imposed for $\phi$, i.e. $\partial_x \phi=0$ if $x$ is the vertical direction.

\begin{figure}[t!]
  %\centering
  \includegraphics[keepaspectratio, width=0.6\textwidth]{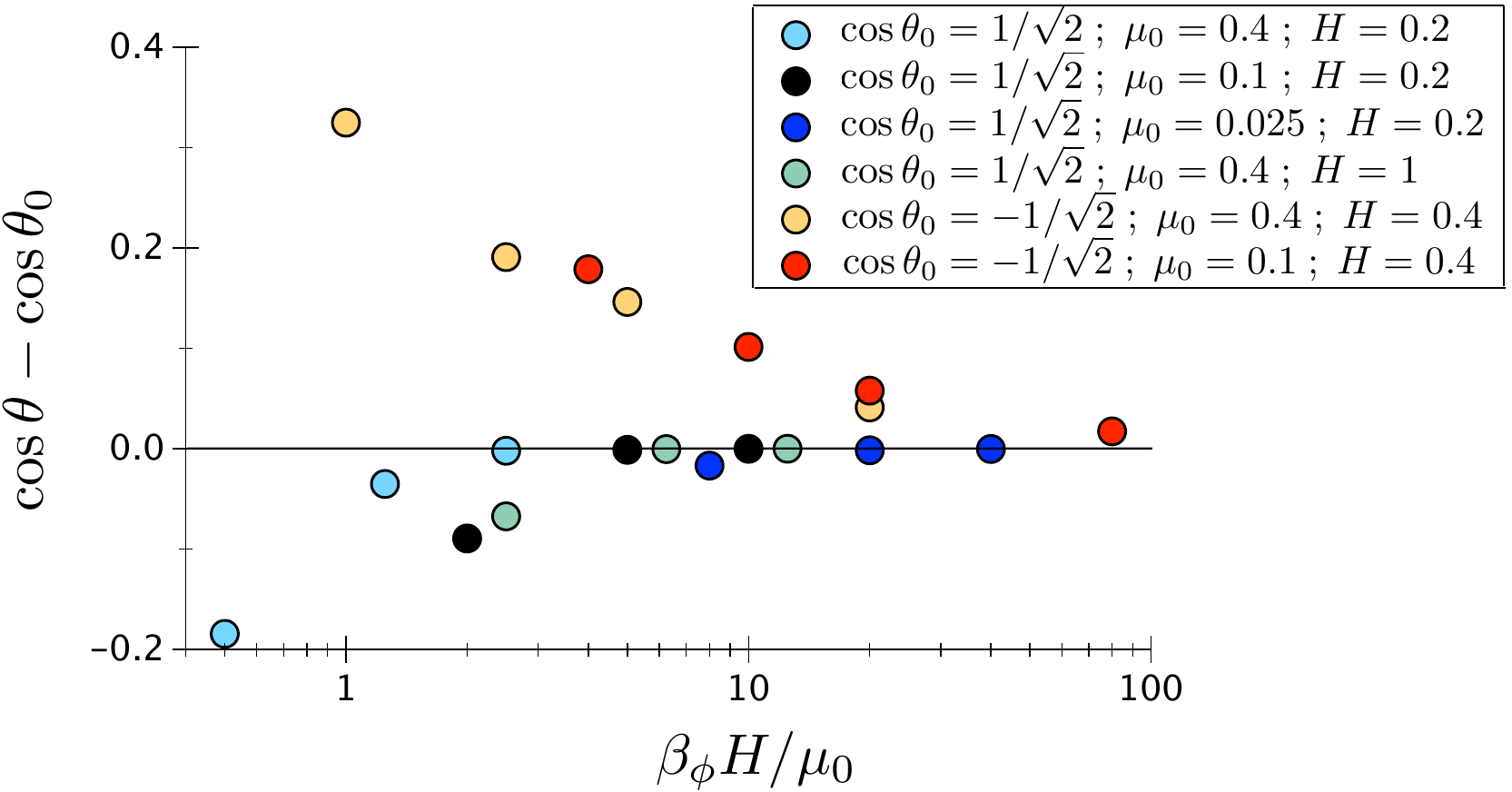}
  \caption{ Convergence  towards $\theta_0$ of the measured angle $\theta$ when $\beta_\phi H/\mu_0$ increases.}
  \label{contact_angle_results}
\end{figure}

In Section \ref{PF_model}, we have introduced in the surface contribution in Eq. (\ref{dot_phi_surface}) the strengthening factor $\beta_\phi$. 
When $\beta_\phi=1$, the SBM formulation corresponds to the one given in Ref. \cite{thornton}. 
We performed simulations with varying $\beta_\phi$, and we found that, depending on $H$, the choice $\beta_\phi=1$ may not allow for an efficient reproduction of $\theta=\theta_0$. 
Indeed, we note from Eqs. (\ref{dot_phi_surface}) and (\ref{dot_phi_bulk}) that a large enough $\mu$ may control the dynamics of $\phi$, leaving a finite $\Phi_{\text{surface}}$. 
We thus plot in Fig. \ref{contact_angle_results} the deviation $\cos \theta - \cos \theta_0$ as a function of $\beta_\phi H/\mu_0$ for varying $H$ and $\mu_0$. 
We see that the convergence of $\theta$ towards $\theta_0$ takes place when $\beta_\phi H/\mu_0 \gg 1$, i.e. for small enough $\mu_0$. One may notice also that this convergence is slower for $\cos \theta_0 = -1/\sqrt{2}$ than for the other case. 

In this test, we have demonstrated the importance of introducing a large enough strengthening factor $\beta_\phi$ ($\beta_\phi \gg \mu_0/H$) in order to achieve a contact angle $\theta=\theta_0$. In opposition, in the previous test for the diffusion equation, the best choice is, according to some tests of ours, not to use any strengthening factor.

\subsection{Traction-free boundary condition for the elastic field} \label{results_elast}

In this part, we set a two-dimensional configuration for the fields $\phi$ and $\mu$, and we solve the elastic problem with traction-free conditions at the P-E interface as explained in Section \ref{model_elast}. 
With SBM we aim at imposing these conditions in the region where $\psi(x,y)$ varies. 
And we perform also a benchmark simulation where we impose explicitly the traction-free conditions at the borders of the simulation domain. 
 
\begin{figure}[t!]
  %\centering
  \includegraphics[keepaspectratio, width=0.5\textwidth]{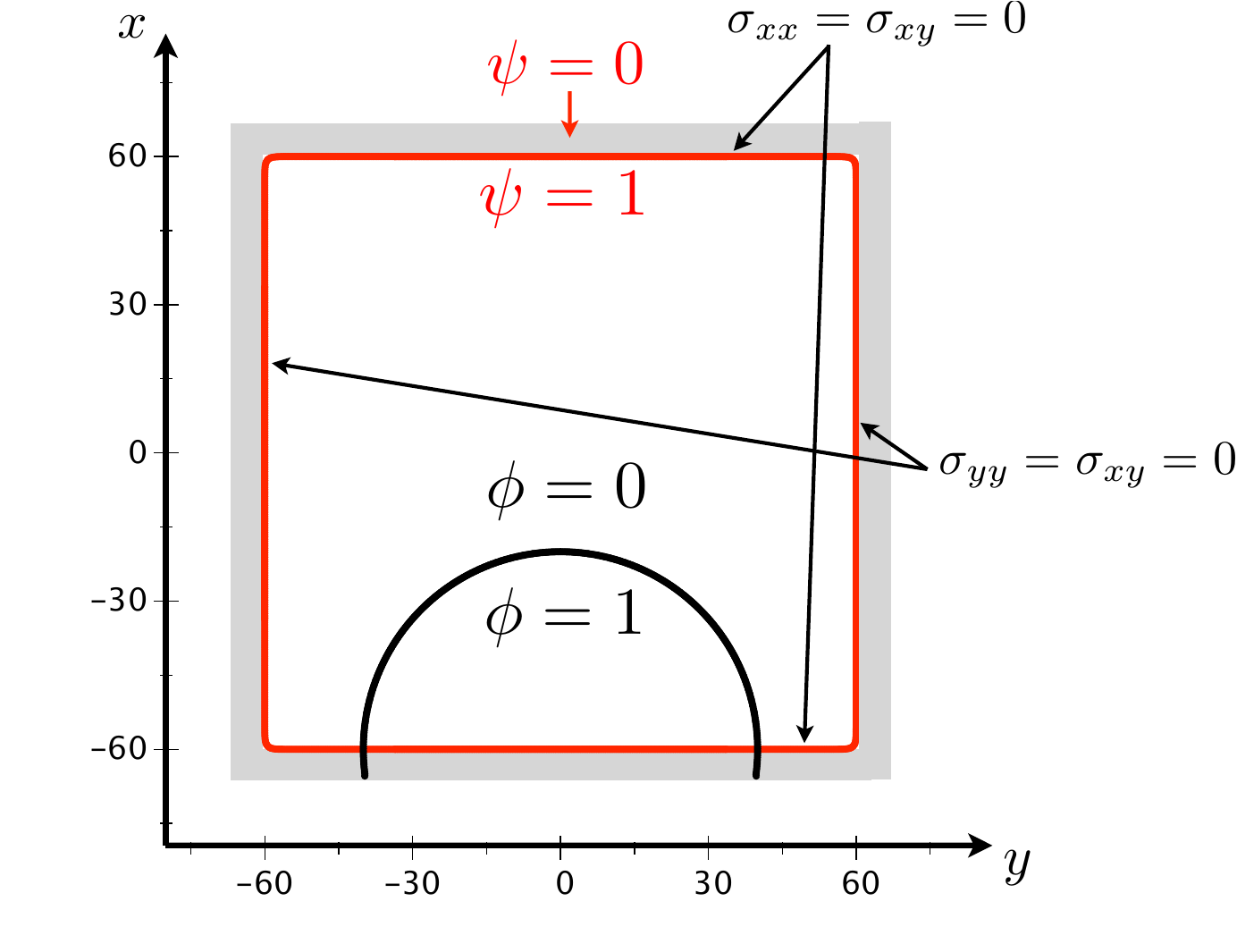}
  \caption{ Configuration of the calculation domain for testing the elastic problem. The benchmark calculation is performed only within the region delimited by the red frame, that represents the P-E interface. The SBM calculation is performed in the whole domain, including the gray region where $\psi=0$. At the P-E interface, traction-free conditions are imposed, whether explicitly in the benchmark calculation or using the surface terms in the equilibrium equation (\ref{elast_eq_SBM}). Elastic deformations result from the coherency assumed at the FP-LFP interface, represented by the black line. }
  \label{elasticity_shape}
\end{figure}

We use a square simulation box with $L=130$ for SBM and $L=120$ for benchmark. For SBM, we use 
\bea
\psi(x,y) = \frac{1}{4} \left[1 - \tanh \frac{|y| - 120}{\sqrt{2} } \right]\left[1 - \tanh \frac{|x| - 120}{\sqrt{2} } \right] .
\eea
The phase field is given by
\bea
\phi(x,y) = \frac{1}{2} \left[  1 - \tanh \left(  \frac{ \sqrt{ (x+60)^2 + y^2 } - 40 }{\sqrt{2} } \right) \right]
\eea
We give a scheme of this configuration Fig. \ref{elasticity_shape}. The red line corresponds to $\psi=0.5$ and the black line corresponds to $\phi=0.5$.
Moreover we set $\mu(x,y)=0$. 
Since $C = \mu + C_{eq}(\phi)$ varies across the FP-LFP interface, spatial variations of the eigenstrain $\varepsilon^0_{ij}(C)$ occur. Owing to the assumption of coherent interface, this yield elastic long-range interactions. 
We search for $u_x$ and $u_y$, the two components of the displacement vector, that correspond to the elastic equilibrium. 
For SBM, they are solutions of Eq. (\ref{elast_eq_SBM}) and we impose irrelevant traction-free conditions at the boundary of the simulation domain where $\psi=0$. 
For the benchmark, we search for the solution of Eq. (\ref{elast_eq_classic}) with traction-free conditions imposed at the simulation domain boundaries (see Fig. \ref{elasticity_shape}). 
In both cases the equilibrium is found with a damped inertial algorithm where the increments of the displacement $\Delta u_i$ and of the velocity $\Delta v_i$ are involved: $\Delta v_i =  A \mathcal F_i - v_i/10$  for SBM and $\Delta  v_i = A \partial_j \sigma_{ij }- v_i/10$ for the benchmark, with $A = 3\times10^{-4}$ and $\Delta u_i = 1.5 \times 10^{-2} v_i$ for both. Hence, when $v_i$ converges to zero, we achieve $\mathcal F_i = 0$ for SBM and $\lambda^0_{ijkl} \partial_j (\varepsilon_{kl} -\varepsilon^0_{kl}) = 0$, i.e. Eq. (\ref{elast_eq_classic}), for the benchmark.

The elastic constants are chosen as:
\bea
\lambda^0_{xxxx} = \lambda^0_{yyyy} = C_{11}=210, \nonumber \\
\lambda^0_{xxyy} = \lambda^0_{yyxx} = C_{12} = 70, \nonumber \\
\lambda^0_{xyxy} = \lambda^0_{yxxy} = \lambda^0_{yxyx} = \lambda^0_{xyyx}=  C_{33} = 70, \label{elast_constants}
\eea 
and 0 otherwise, so that the the crystal is elastically isotropic, i.e. $C_{11} - C_{12} = 2 C_{33}$. These values are typical for a system for which $H \simeq 1$ with a FP-LFP interface free energy of order $10^{-1}$ J/m$^2$, an interface width $W \simeq 10^{-10}$ m and elastic constants of order 100 GPa.
The eigenstrain is  chosen anisotropic for a sake of generality:
\bea
\varepsilon^*_{xx} = 2 \varepsilon^*_{yy} = 2 \times 10^{-2}, \nonumber \\
\varepsilon^*_{xy} = \varepsilon^*_{yx} = 0 . \label{eigenstrains}
\eea

\begin{figure}[t!]
  %\centering
  \includegraphics[keepaspectratio, width=0.5\textwidth]{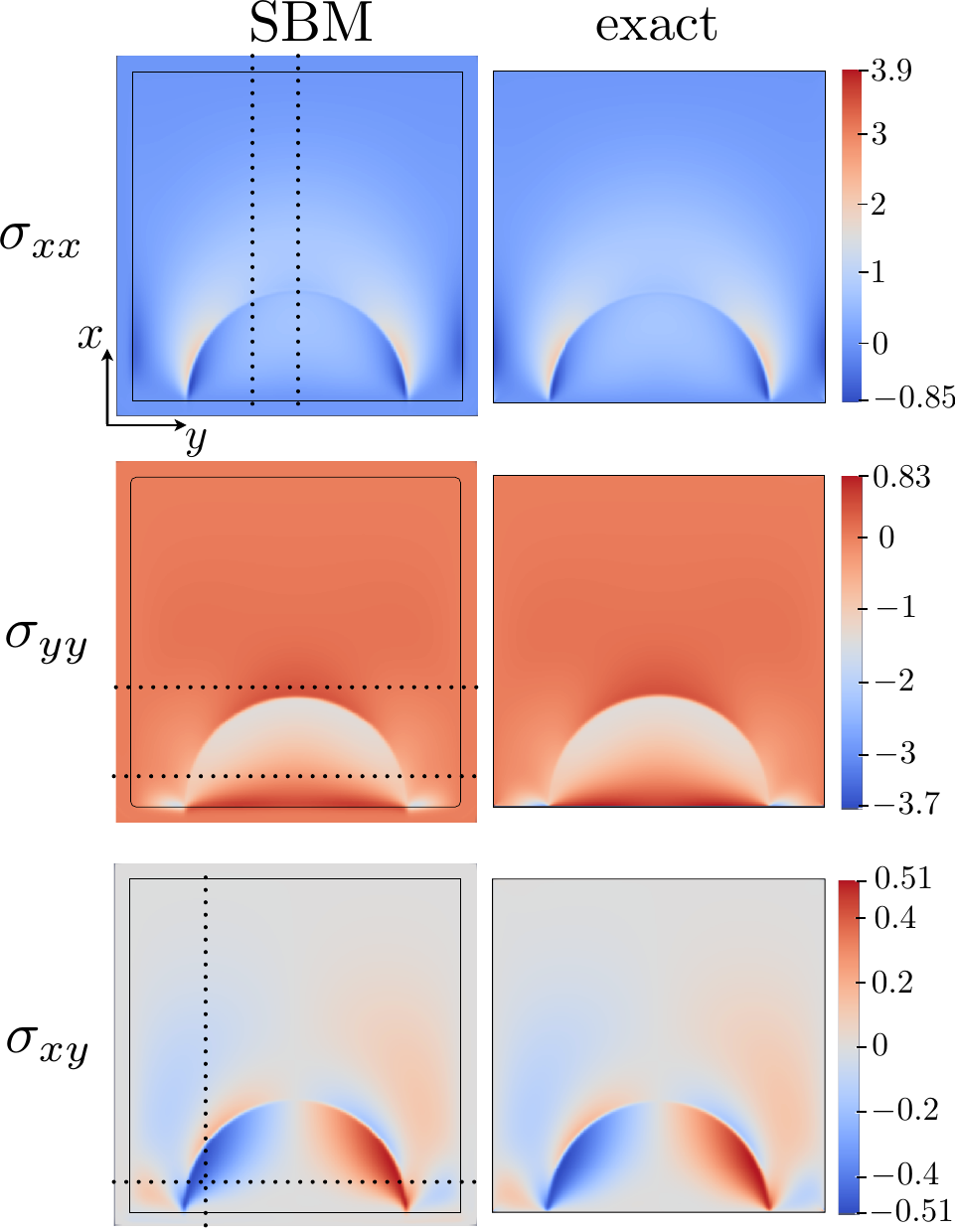}
  \caption{ Color maps, for SBM and the benchmark ("exact"), of the three components $\sigma_{xx}, \sigma_{yy}$ and $\sigma_{xy}$. For SBM, the strengthening factor is $\beta_{\text{el}}=4$. In Figs. \ref{elasticity_sigma_xx}, \ref{elasticity_sigma_yy} and \ref{elasticity_sigma_xy} respectively, plots are presented along the dotted lines. }
  \label{elasticity_images}
\end{figure}

In Fig. \ref{elasticity_images}, we present the color maps for the three components of the stress tensor, and compare the SBM with $\beta_{\text{el}}=4$ and the benchmark ("exact"). Visually, the agreement is very satisfying. However, let us compare SBM and benchmark more precisely. We plot $\sigma_{xx}$, $\sigma_{yy}$ and $\sigma_{xy}$, each of them along the two dotted lines shown in Fig. \ref{elasticity_images}. At the extremity of each of these lines, the corresponding stress should vanish, in virtue of the traction-free condition. 

\begin{figure}[t!]
  %\centering
  \includegraphics[keepaspectratio, width=1\textwidth]{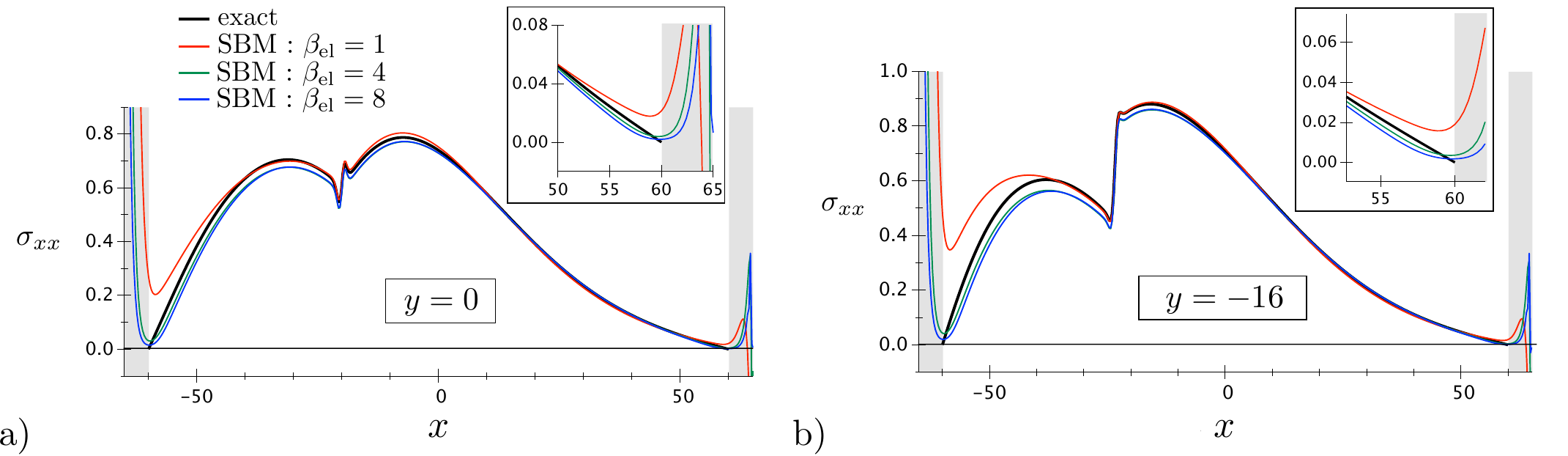}
  \caption{ Value of $\sigma_{xx}$ along the dotted lines in Fig. \ref{elasticity_images} for the different SBM calculations with varying strengthening factor $\beta_{\text{el}}$ and the benchmark ("exact"). Traction-free conditions require $\sigma_{xx} = 0$ at $x = \pm 60$.}
  \label{elasticity_sigma_xx}
\end{figure}

In Figs. \ref{elasticity_sigma_xx}a and \ref{elasticity_sigma_xx}b, we present, for $y=0$ and $y = -16$ respectively, $\sigma_{xx}$ for the benchmark (black) and for the SBM with $\beta_{\text{el}}=1$ (red), $\beta_{\text{el}}=4$ (green) and $\beta_{\text{el}}=8$ (blue). 
We see that the stress is globally faithfully reproduced by the SBM when compared to the benchmark. However, we see that the case $\beta_{\text{el}}=1$ leaves a significant residual $\sigma_{xx}$ in the neighborhood of the P-E interface, about one order of magnitude larger than the residual $\sigma_{xx}$ for $\beta_{\text{el}}=4$ and $\beta_{\text{el}}=8$. While it is clearly visible at $x=-60$, we provide a close up on $x=60$ in the insets. 

%Before analyzing the other components of the stress tensor, let us note that Fig. \ref{elasticity_sigma_xx}a provides an illustration of the necessary continuity of the normal stresses at the FP-LFP interface. Indeed, $\sigma_{xx}$ plotted at $y=0$ represents,  at the interface, one of the two components of the normal stress. We evidence the continuity of $\sigma_{xx}$ in Fig. \ref{elasticity_sigma_xx_zoom_interface} using the close up indicated by the dotted box in Fig. \ref{elasticity_sigma_xx}a. We extrapolate the slopes on both sides of the interface, and we note that their intersection is located at $x=-20W$, i.e. at the center of the P-E interface where $\psi = 0.5$.    

%\begin{figure}[t!]
  %\centering
 % \includegraphics[keepaspectratio, width=0.4\textwidth]{elasticity_sigma_xx_zoom_interface}
  %\caption{ The arrow indicates the progression of time.}
  %\label{elasticity_sigma_xx_zoom_interface}
%\end{figure}

\begin{figure}[t!]
  %\centering
  \includegraphics[keepaspectratio, width=1\textwidth]{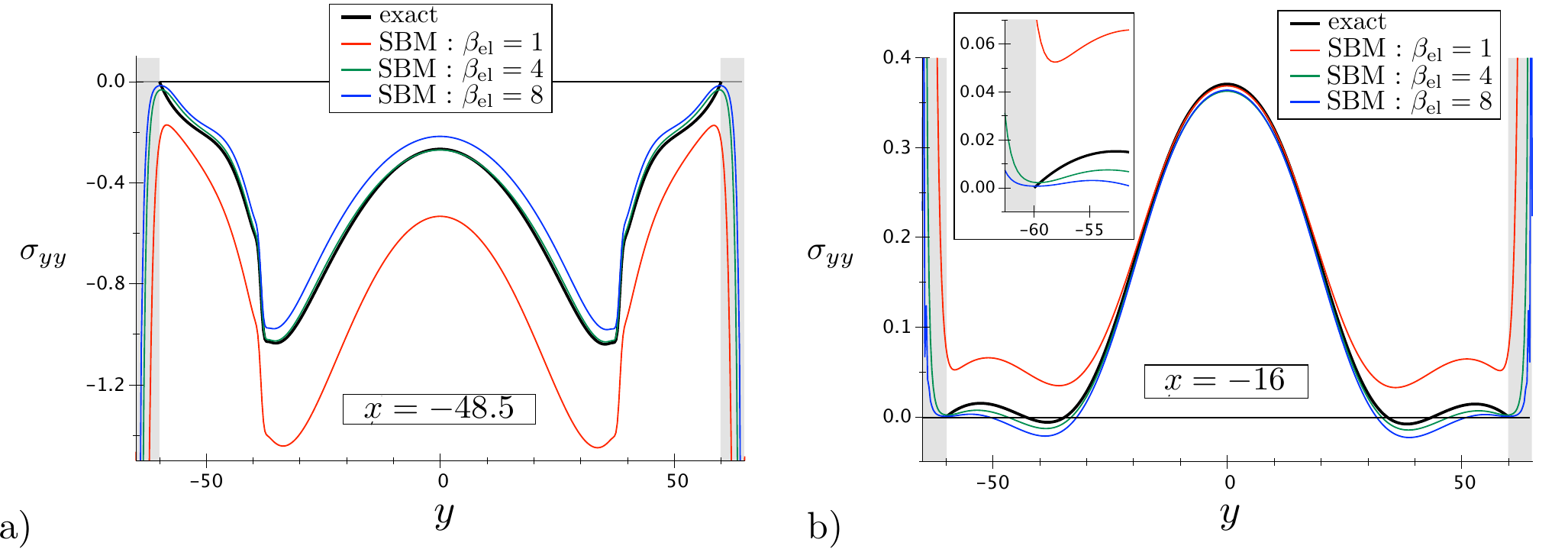}
  \caption{Value of $\sigma_{yy}$ along the dotted lines in Fig. \ref{elasticity_images} for the different SBM calculations with varying strengthening factor $\beta_{\text{el}}$ and the benchmark ("exact"). Traction-free conditions require $\sigma_{yy} = 0$ at $x = \pm 60$.}
  \label{elasticity_sigma_yy}
\end{figure}

We now plot $\sigma_{yy}$ for SBM and benchmark at respectively $x=-48.5$ and $x=-16$ in Figs. \ref{elasticity_sigma_yy}a and \ref{elasticity_sigma_yy}b. 
We see that $\beta_{\text{el}}=1$ (red) yields a rather large error when compared to the benchmark. 
In particular, the magnitude of the stress exhibits a deviation from benchmark that is not restricted to the neighborhood of the P-E interface. For example, at the center of the inclusion, i.e. at $y=0$ in Fig. \ref{elasticity_sigma_yy}a, the compressive stress is overestimated by a factor 2. And we identify again, as illustrated in the inset in Fig. \ref{elasticity_sigma_yy}b, a significant residual stress at the P-E interface. 
We also note that the best agreement with the benchmark is obtained for $\beta_{\text{el}}=4$ (green).

\begin{figure}[t!]
  %\centering
  \includegraphics[keepaspectratio, width=1\textwidth]{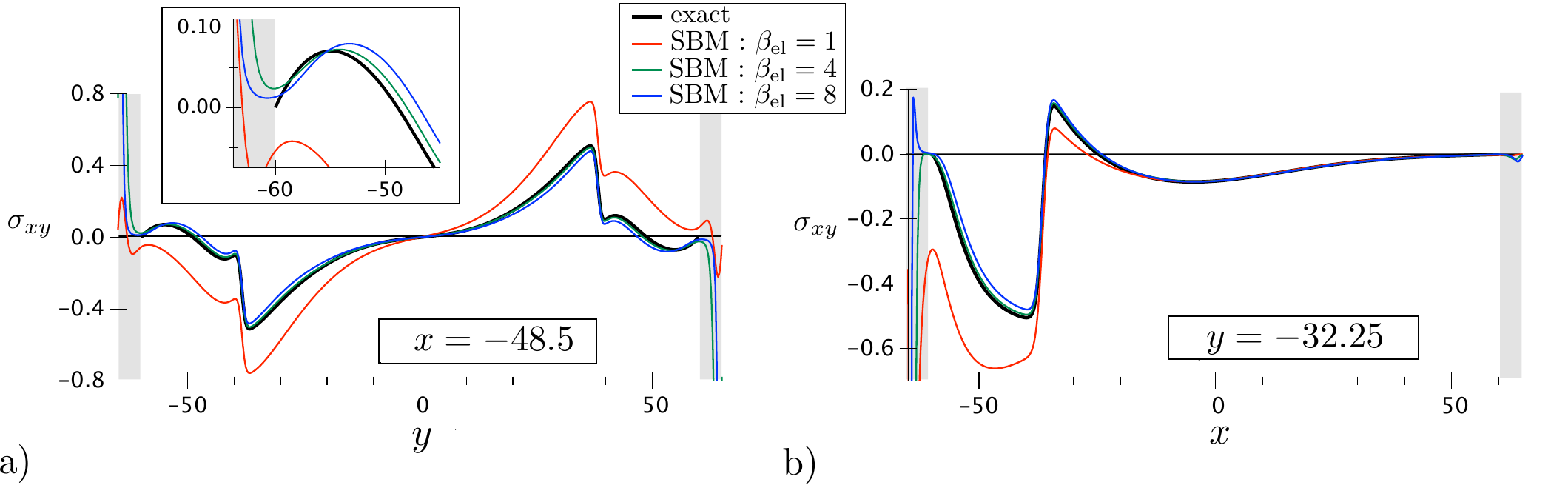}
  \caption{Value of $\sigma_{xy}$ along the dotted lines in Fig. \ref{elasticity_images} for the different SBM calculations with varying strengthening factor $\beta_{\text{el}}$ and the benchmark ("exact"). Traction-free conditions require $\sigma_{xy} = 0$ at $x = \pm 60$.}
  \label{elasticity_sigma_xy}
\end{figure}

Before commenting this, let us analyze the shear stress. In Figs. \ref{elasticity_sigma_xy}a and \ref{elasticity_sigma_xy}b, we present $\sigma_{xy}$ along, respectively, $x=-48.5W$ and $y=-32.25W$. As for $\sigma_{yy}$, we observe severe deviations for $\beta_{\text{el}}=1$. And here also, we note that the best agreement between SBM and benchmark is provided by $\beta_{\text{el}}=4$. With the help of the inset in Fig. \ref{elasticity_sigma_xy}a, we are able to identify the mechanism responsible for the existence of an optimum value for $\beta_{\text{el}}$. Indeed, we see that the larger $\beta_{\text{el}}$ is, the closer to 0 but also the more flat $\sigma_{xy}$ is at the P-E interface. The region within which the condition $\sigma_{xy}=0$ holds becomes wider. This may also be observed in Figs. \ref{elasticity_sigma_xx} and \ref{elasticity_sigma_yy}.   
Hence, since the equilibration of mechanical forces involves long range interactions, the level of stress in the whole sample may be influenced by $\beta_{\text{el}}$. As an illustration, we clearly see a systematic deviation of $\sigma_{yy}$ from the benchmark  for $\beta_{\text{el}}=8$ in Fig. \ref{elasticity_sigma_yy}a.
\\

{\bf Discussion.}
We have demonstrated the necessity of introducing the strengthening factor $\beta_{\text{el}}$ in order to achieve a correct level of stress within the particle and a traction-free P-E interface. This takes a special meaning in the context of Li-ion batteries. 
Indeed, one of the governing factor of the lithiation/delithiation dynamics is the reaction (insertion/extraction) rate at the P-E interface, given by Eq. (\ref{reaction_rate}). And in the latter, the overpotential $\Delta$  depends on $\delta F/\delta C$ [Eq. (\ref{Delta})], in which contribute the chemical potential $\mu$ but also the elastic field via $-\sigma_{ij} \varepsilon^*_{ij}$ [see Eq. (\ref{def_df_dc})]. 

\begin{figure}[t!]
  %\centering
  \includegraphics[keepaspectratio, width=1\textwidth]{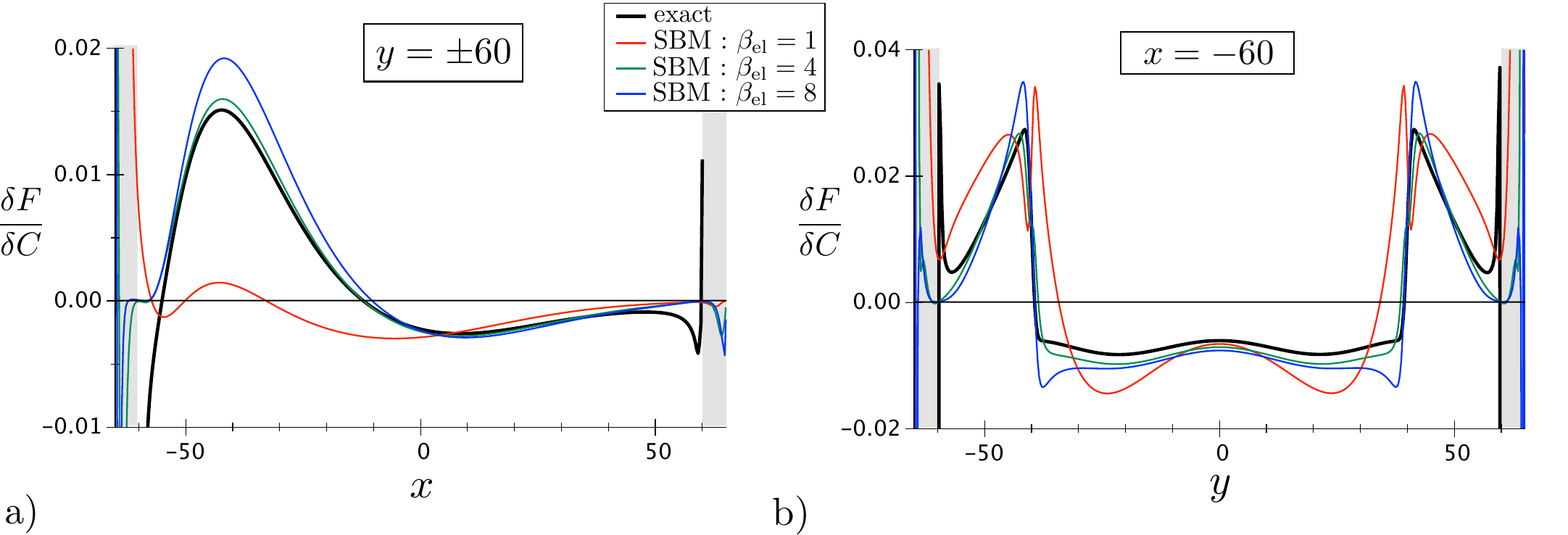}
  \caption{Elastic contribution to the chemical potential (here $\delta F/\delta C = - \sigma_{ij} \varepsilon^*_{ij}$) at the bottom ($x=-60$) and lateral ($y=\pm60$) P-E interfaces, showing that the SBM fails at reproducing the benchmark when $\beta_{\text{el}} = 1$.  }
  \label{elasticity_reaction_rate}
\end{figure}

Therefore, the value of $\delta F/\delta C$ at the P-E interface is an important ingredient of the dynamics that has to be correctly reproduced by the SBM. We plot $\delta F/\delta C$ (here $\delta F/\delta C = -\sigma_{ij} \varepsilon^*_{ij}$ since $\mu=0$) along the two main P-E interfaces $x=-60W$ and $y=\pm60W$ (at the third $x=60W$, the elastic stress is almost null)  in Figs. \ref{elasticity_reaction_rate}a and \ref{elasticity_reaction_rate}b respectively.
  
First, one should note that for the benchmark, an artifact appears. The extremities of the two curves correspond to  corners of our simulation domain, and in the benchmark simulation, the grid points corresponding to a corner are problematic. 
Indeed, among their first-neighbors, two of them need to be determined via BCs. In opposition, only one first-neighbor needs to be determined for the other grid points on the boundary (where the two traction-free conditions fix the two displacement components). Moreover, due to the second derivatives $\partial_{xy} u_i$ in the elastic equilibrium equations, a second-neighbor of the corner grid points needs also to be determined by BCs. We clearly lack equations to apply appropriate conditions at the corner grid points. Nevertheless, this deficit does not influence the elastic field on the long range, as can be seen in the color maps in Fig. \ref{elasticity_images} (one may perform a deep zoom and distinguish the small region where the improper BCs play a role). 
We know that $\delta F/\delta C$ should vanish at $x= \pm 60W$ in Fig. \ref{elasticity_reaction_rate}a and at $y= \pm 60W$ in Fig. \ref{elasticity_reaction_rate}b, and this is faithfully reproduced by our SBM simulations with $\beta_{\text{el}}=4$ and $\beta_{\text{el}}=8$. We see also that, again, $\beta_{\text{el}}=4$ provides the least deviation from the benchmark. 

According to Fig. \ref{elasticity_reaction_rate}, the elastic contribution to the overpotential may differ strongly, depending on $\beta_{\text{el}}$. 
For example, along the $x$-axis in Fig. \ref{elasticity_reaction_rate}a, $\delta F/\delta C$ presents a pronounced peak of magnitude close to $0.015$ for $\beta_{\text{el}}=4$ and the benchmark (and an even more pronounced peak for $\beta_{\text{el}}=8$ which we consider overestimated). 
This is illustrated by $\sigma_{xx}$ in Fig. \ref{elasticity_images} with the darker blue region on the lateral boundaries of the simulation domain.
Thus, for a situation where $|\mu| \simeq 0.015$, the overpotential driving the electrochemical reaction would be significantly affected by elastic effects.
In opposition, when the SBM calculation is performed with $\beta_{\text{el}}=1$ (red), we see in Fig. \ref{elasticity_reaction_rate}a that the peak is almost inexistent, meaning that the overpotential would remain mostly unaffected by elastic effects. 
This is a qualitative difference that illustrates how the interplay between the BCs operates. An appropriate choice for the strengthening factor $\beta_{\text{el}}$ enabling the traction-free P-E interface reveals crucial for achieving the correct insertion rate. 
%Thus the accuracy that the model allows may have large consequences on the description of the charge/discharge cycles.  

\section{Potentio-static simulations}

In order to illustrate the applicability of our approach, we finally present the two-dimensional  lithiation of a rectangular particle under potentio-static conditions, for which elastic deformations are accounted for. 
This represents a full coupling of the different physical processes described in the previous sections. 
We also perform the simulation without elasticity for comparison.
We use the same elastic constants and eigenstrains as in the previous section, given by Eqs. (\ref{elast_constants}) and (\ref{eigenstrains}). 
We take $H=0.2$,  and we use $\cos \theta_0 = 0$ with $\beta_{\phi}=10$. We choose an isotropic diffusion with $D_x^0 = D_y^0 = D$. We choose also to neglect the difference of open-circuit potential between FP and LFP, i.e. $\mu_0^{LFP} = \mu_0^{FP} = \mu_0^{\text{ref}}$. 

\begin{figure}[t!]
  %\centering
  \includegraphics[keepaspectratio, width=0.95\textwidth]{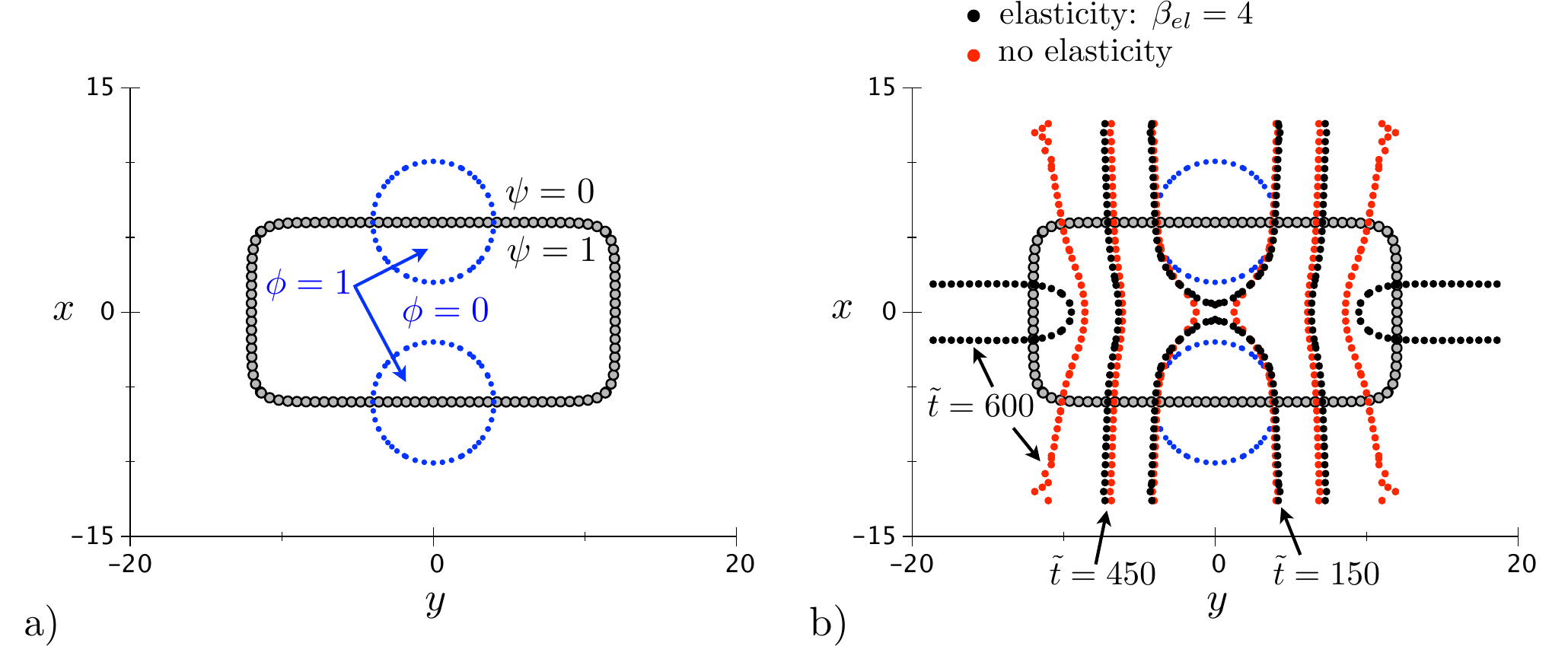}
  \caption{ (a) Initial configuration for the potentio-static SBM simulation. The particle, where $\psi=1$, is a rectangle with a width $\rho_x = 6$ and a length $\rho_y = 12$. Two LFP seeds, where $\phi=1$, are introduced. (b) Evolution driven by a constant $\mu_0$ of the FP-LFP interface, whose location is given at different times $\tilde t = Dt/W^2$, with ($\beta_{\text{el}}=4$) and without elasticity (see text for further information).}
  \label{simul_shape}
\end{figure}

The rectangular particle is described by 
\bea
\psi(x,y) = \frac{1}{4} \left[ 1 - \tanh \left( \frac{|x| - \rho_x}{\sqrt{2} } \right) \right] \left[ 1 - \tanh \left( \frac{|y| - \rho_y}{\sqrt{2} } \right) \right],
\eea 
where $\rho_x = \rho_y/2 = 6$, yielding an area $\m A = \int_S \psi \; dS \simeq 4 \rho_x \rho_y$, where $dS = dx \; dy$. 
As shown in Fig. \ref{simul_shape}a, we initially introduce two circular LFP seeds, i.e. 
 \bea
 \phi (x,y) = 1- \frac{1}{2} \left[  \tanh \left( \frac{\sqrt{(x+\rho)^2 + y^2} - r}{\sqrt{2} } \right) + \tanh \left( \frac{\sqrt{(x-\rho)^2 + y^2} - r}{\sqrt{2} } \right) \right] ,
 \eea
 with a radius $r=4$. This value of $r$ ensures that the two hyperbolic tangents do not overlap.

\begin{figure}[t!]
  %\centering
  \includegraphics[keepaspectratio, width=0.86\textwidth]{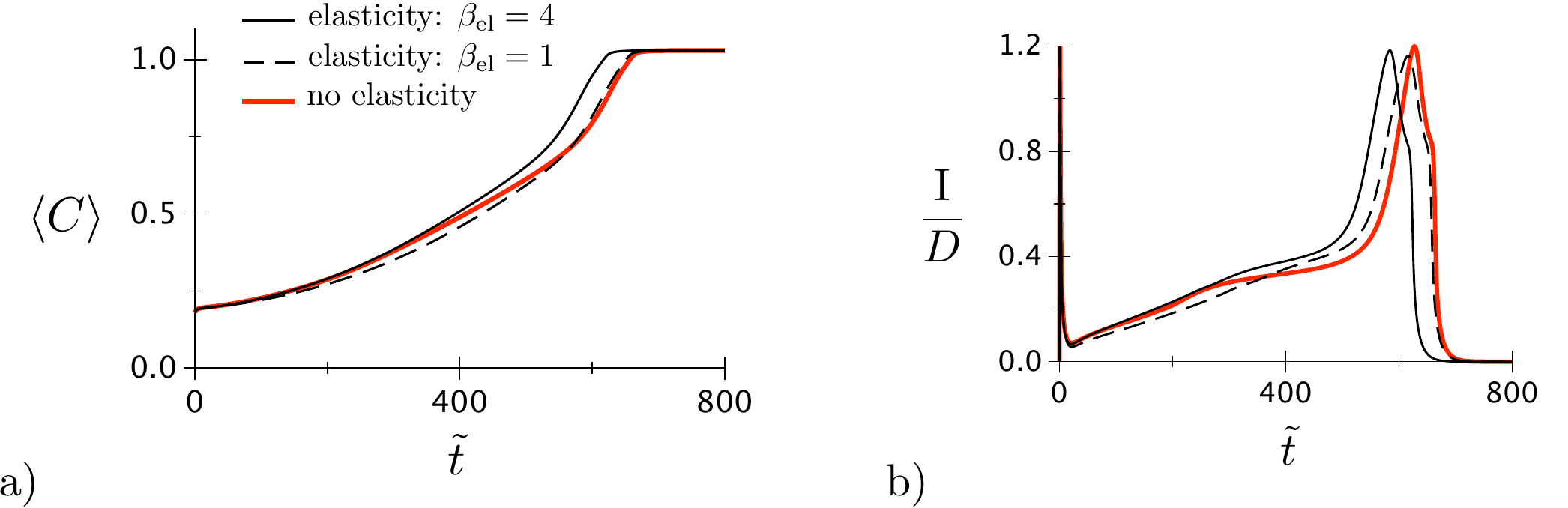}
  \caption{Potentio-static SBM simulation. With ($\beta_{\text{el}}=1$ and $\beta_{\text{el}}=4$) and without elastic effects, are presented the evolutions as a function of $\tilde t$ of : (a) the state of charge $\langle C \rangle$; (b) the dimensionless electrical current ${\text I}/D$.}
  \label{simul_conc}
\end{figure}

We impose a constant external potential (potentio-static conditions) $\mu_0 = \mu_0^0 - \mu_0^{\text{ref}} = 0.03$, and initially we set $\mu=0$ in the simulation domain. 
The lithiation is thus initiated with a state of charge $\langle C \rangle = \pi r^2/\m A \simeq 0.17$.
As previously we assume a linear insertion rate ${\text J}_{\text{ins}} = (D /l_{\text{ins}}) (\mu_0 - \delta F/\delta C)$, with here $l_{\text{ins}} = 1$. 
In Fig. \ref{simul_shape}b, we present, for $\tilde t = Dt/W^2 = 150, 450$ and 600, the position of the FP-LFP interfaces, i.e. the locus of $\phi=0.5$. In black, elastic deformations are present and the strengthening factor is $\beta_{\text{el}} = 4$, while in red elastic deformation are absent.
In both cases, the two LFP seeds merge. Then, the FP-LFP interface propagates laterally and subsequently engulfs  the FP phase, that finally disappears.  
However, as can be seen in Figs. \ref{simul_conc}a and \ref{simul_conc}b where we plot respectively, as a function of $\tilde t$, the state of charge $\langle C \rangle$ and the corresponding dimensionless electrical current  $\text{I}/D = (\m A/D) \dot{ \langle  C \rangle}$ [see Eqs. (\ref{state_of_charge}) and (\ref{relation_i})], the dynamics is accelerated in presence of elastic deformation (note that in two dimensions the current $\text{I}$ is measured in m$^2$s$^{-1}$ alike $D$, rendering $\text{I}/D$ dimensionless). 
The state of charge reaches its asymptotic value $\langle C \rangle = C_{eq}(\phi=1) + \mu_0 = 1.03$ faster, and the final increase in electrical current corresponding to the vanishing of the FP phase takes place earlier. 
This is rationalized physically by the following argument. While the FP-LFP interface propagates laterally, elastic deformations develop in the FP phase in the form of a stretching of the crystalline lattice, i.e. yielding a positive stress $\sigma_{ij}$. The  overpotential $\Delta$ at the P-E interface therefore increases [see the short discussion following Eq. (\ref{Delta})] and this favors the Li insertion and accelerates the lithiation of the particle.

One should note that, here also, the introduction of the strengthening factor $\beta_{\text{el}}$ reveals crucial. Indeed, as can be seen in Fig. \ref{simul_conc}, the case $\beta_{\text{el}}=1$ (dashed line) does not provide the same lithiation dynamics as  $\beta_{\text{el}}=4$. In comparison, for $\beta_{\text{el}}=8$, one obtains curves that are undistinguishable from the case $\beta_{\text{el}}=4$.

\section{Conclusion}

In this article, we present a phase field model to tackle the problem of biphasic lithiation/delithiation dynamics in Li-ion batteries.  
The model is based on the Allen-Cahn approach with a conserved field and a non-conserved field, supplemented by the elastic displacements. 
The model uses the Smoothed Boundary Method (SBM), that defines a smoothly varying indicator field discriminating the cathode particle and the surrounding liquid electrolyte, in order to achieve appropriate boundary conditions (BCs) at the particle-electrolyte (P-E) interface. 
The procedure relies on introducing surface terms in the evolution equations obeyed by the different fields.  
After presenting the model, we investigate the capabilities of the SBM for the three types of BCs that have to be imposed at the P-E interface: the electro-chemical reaction rate describing the Li insertion/extraction, the contact angle at the position where the P-E interface meets the interface between the two phases, and the traction-free conditions for the elastic field. 
For each of them, we perform calculations in an ideal situation where only the physical process that plays a role is activated (for example we exclude diffusion and elasticity when studying the contact angle). The results are compared with those of  benchmark calculations where the BCs are explicitly imposed at the boundary of the calculation domain.
We find that, in order to impose the correct contact angle and to achieve correct levels of elastic stress with traction-free conditions, one needs to introduce strengthening factors in the surface terms on which rely the SBM.
This reveals crucial for a correct prediction of the overpotential at the P-E interface that drives insertion/extraction of Li.   
The importance of the strengthening factor is confirmed by the simulation of a lithiation dynamics under potentio-static conditions where all physical processes are at play. 
The latter simulation, that in particular evidences the influence of elastic effects, is performed to illustrate the applicability of our approach, and shows that our work opens the way for a phase field modeling of the lithiation/delithiation dynamics of arbitrarily shaped cathode particles using the Smoothed Boundary Method.

\section{Acknowledgements} 

We thank Shahed Rezaei for fruitfull discussions. Two of the authors (A.Y. and A.D.) also want to thank the French network on electrochemical energy storage (RS2E) and the French region Hauts-de-France for the financial support.

%\section{Acknowledgements}
%
%This study was funded by the Deutsche Forschungsgemeinschaft (DFG, German Research Foundation) under grant number AP196/17-1. 
%
%
%

\end{document}